\documentstyle[twoside]{article}
\catcode`\@=11
\long\def\@makefntext#1{
\protect\noindent \hbox to 3.2pt {\hskip-.9pt  
$^{{\eightrm\@thefnmark}}$\hfil}#1\hfill}

\def\@makefnmark{\hbox to 0pt{$^{\@thefnmark}$\hss}}
\def\ps@myheadings{\let\@mkboth\@gobbletwo
\def\@oddhead{\hbox{}
\rightmark\hfil\eightrm\thepage}   
\def\@oddfoot{}\def\@evenhead{\eightrm\thepage\hfil
\leftmark\hbox{}}\def\@evenfoot{}
\def\sectionmark##1{}\def\subsectionmark##1{}}
\oddsidemargin=\evensidemargin
\newcounter{sectionc}\newcounter{subsectionc}\newcounter{subsubsectionc}
\renewcommand{\section}[1] {\vspace{12pt}\addtocounter{sectionc}{1} 
\setcounter{subsectionc}{0}\setcounter{subsubsectionc}{0}\noindent 
 {\tenbf\thesectionc. #1}\par\vspace{5pt}}
\renewcommand{\subsection}[1]
{\vspace{12pt}\addtocounter{subsectionc}{1} 
 \setcounter{subsubsectionc}{0}\noindent 
{\bf\thesectionc.\thesubsectionc. {\kern1pt \bfit #1}}\par\vspace{5pt}}
\renewcommand{\subsubsection}[1]
{\vspace{12pt}\addtocounter{subsubsectionc}{1}
 \noindent{\tenrm\thesectionc.\thesubsectionc.\thesubsubsectionc.
 {\kern1pt \tenit #1}}\par\vspace{5pt}}

\newcounter{appendixc}
\newcounter{subappendixc}[appendixc]
\newcounter{subsubappendixc}[subappendixc]
\renewcommand{\thesubappendixc}{\Alph{appendixc}.\arabic{subappendixc}}
\renewcommand{\thesubsubappendixc}
 {\Alph{appendixc}.\arabic{subappendixc}.\arabic{subsubappendixc}}
\renewcommand{\appendix}[1] {\vspace{12pt}
        \refstepcounter{appendixc}
        \setcounter{figure}{0}
        \setcounter{table}{0}
        \setcounter{lemma}{0}
        \setcounter{theorem}{0}
        \setcounter{corollary}{0}
        \setcounter{definition}{0}
        \setcounter{equation}{0}
        \renewcommand{\thefigure}{\Alph{appendixc}.\arabic{figure}}
        \renewcommand{\thetable}{\Alph{appendixc}.\arabic{table}}
        \renewcommand{\theappendixc}{\Alph{appendixc}}
        \renewcommand{\thelemma}{\Alph{appendixc}.\arabic{lemma}}
        \renewcommand{\thetheorem}{\Alph{appendixc}.\arabic{theorem}}
     \renewcommand{\thedefinition}{\Alph{appendixc}.\arabic{definition}}
       \renewcommand{\thecorollary}{\Alph{appendixc}.\arabic{corollary}}
        \renewcommand{\theequation}{\Alph{appendixc}.\arabic{equation}}
        \noindent{\tenbf Appendix \theappendixc #1}\par\vspace{5pt}}
\newcommand{\subappendix}[1] {\vspace{12pt}
        \refstepcounter{subappendixc}
        \noindent{\bf Appendix \thesubappendixc. {\kern1pt \bfit #1}}
 \par\vspace{5pt}}
\newcommand{\subsubappendix}[1] {\vspace{12pt}
        \refstepcounter{subsubappendixc}
       \noindent{\rm Appendix \thesubsubappendixc. {\kern1pt \tenit #1}}
 \par\vspace{5pt}}
\newcommand{\textlineskip}{\baselineskip=13pt}
\newcommand{\smalllineskip}{\baselineskip=10pt}
\def\eightcirc{
\begin{picture}(0,0)
\put(4.4,1.8){\circle{6.5}}
\end{picture}}
\def\eightcopyright{\eightcirc\kern2.7pt\hbox{\eightrm c}} 
\newcommand{\copyrightheading}[1]
 {\vspace*{-2.5cm}\smalllineskip{\flushleft
 {\footnotesize International Journal of Modern Physics B, #1}\\
 {\footnotesize $\eightcopyright$\, World Scientific Publishing
  Company}\\
  }}

\newcommand{\publisher}[2]{{\begin{center}\footnotesize\smalllineskip 
 Received #1
 \end{center}
 }}
\def\abstracts#1#2#3{{
 \centering{\begin{minipage}{4.5in}\baselineskip=10pt\footnotesize
 \parindent=0pt #1\par 
 \parindent=15pt #2\par
 \parindent=15pt #3
 \end{minipage}}\par}}

\renewenvironment{thebibliography}[1]   
 {\frenchspacing
  \ninerm\baselineskip=11pt
  \begin{list}{\arabic{enumi}.}
 {\usecounter{enumi}\setlength{\parsep}{0pt}
  \setlength{\leftmargin 12.7pt}{\rightmargin 0pt} 
  \setlength{\itemsep}{0pt} \settowidth
 {\labelwidth}{#1.}\sloppy}}{\end{list}}
\newcounter{itemlistc}
\newcounter{romanlistc}
\newcounter{alphlistc}
\newcounter{arabiclistc}

\newcommand{\fcaption}[1]{
        \refstepcounter{figure}
        \setbox\@tempboxa = \hbox{\footnotesize Fig.~\thefigure. #1}
        \ifdim \wd\@tempboxa > 5in
           {\begin{center}
        \parbox{5in}{\footnotesize\smalllineskip Fig.~\thefigure. #1}
            \end{center}}
        \else
             {\begin{center}
             {\footnotesize Fig.~\thefigure. #1}
              \end{center}}
        \fi}
\newcommand{\tcaption}[1]{
        \refstepcounter{table}
        \setbox\@tempboxa = \hbox{\footnotesize Table~\thetable. #1}
        \ifdim \wd\@tempboxa > 5in
           {\begin{center}
        \parbox{5in}{\footnotesize\smalllineskip Table~\thetable. #1}
            \end{center}}
        \else
             {\begin{center}
             {\footnotesize Table~\thetable. #1}
              \end{center}}
        \fi}
\def\@citex[#1]#2{\if@filesw\immediate\write\@auxout
 {\string\citation{#2}}\fi
\def\@citea{}\@cite{\@for\@citeb:=#2\do
 {\@citea\def\@citea{,}\@ifundefined
 {b@\@citeb}{{\bf ?}\@warning
 {Citation `\@citeb' on page \thepage \space undefined}}
 {\csname b@\@citeb\endcsname}}}{#1}}
\newif\if@cghi
\def\cite{\@cghitrue\@ifnextchar [{\@tempswatrue
 \@citex}{\@tempswafalse\@citex[]}}
\def\citelow{\@cghifalse\@ifnextchar [{\@tempswatrue
 \@citex}{\@tempswafalse\@citex[]}}
\def\@cite#1#2{{$\null^{#1}$\if@tempswa\typeout
 {IJCGA warning: optional citation argument 
 ignored: `#2'} \fi}}

\def\pmb#1{\setbox0=\hbox{#1}
 \kern-.025em\copy0\kern-\wd0
 \kern.05em\copy0\kern-\wd0
 \kern-.025em\raise.0433em\box0}

\def\fnt#1#2{\footnotetext{\kern-.3em
 {$^{\mbox{\scriptsize #1}}$}{#2}}}
\def\fpage#1{\begingroup
\voffset=.3in
\thispagestyle{empty}\begin{table}[b]\centerline{\footnotesize #1}
 \end{table}\endgroup}
\def\runninghead#1#2{\pagestyle{myheadings}
\markboth{{\protect\footnotesize\it{\quad #1}}\hfill}
{\hfill{\protect\footnotesize\it{#2\quad}}}}
\font\tenrm=cmr10
\font\tenit=cmti10 
\font\tenbf=cmbx10
\font\bfit=cmbxti10 at 10pt
\font\ninerm=cmr9
\font\nineit=cmti9
\font\ninebf=cmbx9
\font\eightrm=cmr8

\def\qed{\hbox{${\vcenter{\vbox{   
   \hrule height 0.4pt\hbox{\vrule width 0.4pt height 6pt
   \kern5pt\vrule width 0.4pt}\hrule height 0.4pt}}}$}}

\def\bsc{{\sc a\kern-6.4pt\sc a\kern-6.4pt\sc a}} 
\def\bflatex{\bf L\kern-.30em\raise.3ex\hbox{\bsc}\kern-.14em 
T\kern-.1667em\lower.7ex\hbox{E}\kern-.125em X}

\input epsf

\begin{document}
\runninghead{Z. Toroczkai}{The Brownian vacancy driven walk}

\normalsize\textlineskip
\thispagestyle{empty}
\setcounter{page}{1}

\copyrightheading{}  

\vspace*{0.88truein}

\fpage{1}
\centerline{\bf THE BROWNIAN VACANCY DRIVEN WALK}
\vspace*{0.37truein}
\centerline{\footnotesize ZOLT\'AN TOROCZKAI}
\vspace*{0.018truein}
\centerline{\footnotesize\it Center for Stochastic Processes
in Science and Engineering, and }
\baselineskip=10pt
\centerline{\footnotesize\it
Department of Physics,
Virginia Polytechnic Institute and State University,}
\baselineskip=10pt
\centerline{\footnotesize\it
 Blacksburg,
Virginia 24061-0435, USA} 

\vspace*{0.225truein}
\publisher{27 May 1997}{ }

\vspace*{0.21truein}

\abstracts{ 
We investigate the lattice walk performed by
a tagged member of an infinite
`sea' of particles filling a d-dimensional
lattice, in the presence of a Brownian vacancy. 
Particle-particle exchange is forbidden; 
the only interaction between them
being hard core exclusion. The 
tagged particle, differing from the
others only by its tag, moves only 
when it exchanges places with the hole.
In this sense, it is a lattice walk 
``driven'' by the Brownian vacancy. The
probability distributions for its 
displacement and for the number of steps
taken, after $n$-steps of the vacancy, 
are derived. Surprisingly, none of them is a Gaussian!
It is shown that the only nontrivial 
 dimension where the walk is recurrent
is $d=2$.}{}{}

\vspace*{1pt}\textlineskip 

\section{Introduction}

In solids, long-range diffusion of 
atoms mediated by vacancies 
(the so-called vacancy mechanism) is an  
ubiquitous phenomenon\cite{Sholl,Alexander1,Beijeren,Kehr,Palmer}. 
An atom 
can perform a  
displacement only if there is a neighbouring 
vacancy. For low enough vacancy  
densities the motion performed by each 
of these vacancies can be considered  
as a simple (Brownian) random walk on 
the crystalline lattice, independent 
of the other walkers.  Let us tag one of the atoms.
 Due to encounters with the  
vacancies, this atom performs some kind 
of random walk. One question  
naturally arises: what are the properties 
of this walk? Specifically, we may  
ask: What is the probability distribution 
to find this atom at a certain  
site? How long does it take, on the average, 
to get there, etc? One central  
result of this study is that the probability
 distributions {\em are not}  
Gaussians, being very differrent from the 
walk performed by the vacancy.  
Since this walk is ``driven'' by a 
Brownian hole, we introduced a name for  
it: ``the Brownian vacancy driven walk''.  
  
The precise formulation of 
this problem is stated in  
Section 2 together with the 
derivation of the 
complete probability
distribution characterizing 
the couple tag-vacancy. 
In Section 3 we cover some 
important relations from the  
theory of lattice walks and 
walks with taboo sites, so 
that in Section 4, we can 
explicitely present the  
probability distributions for the 
displacement and for 
the number of steps  
the tagged particle made. It is 
also shown that the only non-trivial  
dimension where the walk 
is recurrent is $d=2$. 

Before we present our analysis, we remark that, 
after completion of our work, we were informed
(Krapivsky) of the study by 
Brummelhuis and Hilhorst\cite{Brummelhuis}.
These authors considered the same problem, but only
in two dimensions. Our approach differs from theirs,
 in that we considered systems  
in arbitrary dimensions: $d$. By leaving 
$d$ as a {\em parameter} in the  
theory, we see the uniqueness of $d=2$ 
systems within this wider context.  
Further, we have analyzed several other
 quantities absent in Ref. 17. 
 See, for example, Section 4.1.  
 Of course, where we do have overlap,
 the results of 17 are  
reproduced exactly.

\section{Formulation of the problem
and construction of the probability  distribution}  
  
Consider a $d$-dimensional hypercubic, 
infinite lattice. Assume, that every  
lattice site is occupied by 
\begin{figure}[htbp]
\vspace*{-6pt} 
\hspace*{2cm}
\epsfxsize = 3.4 in 
\epsfbox{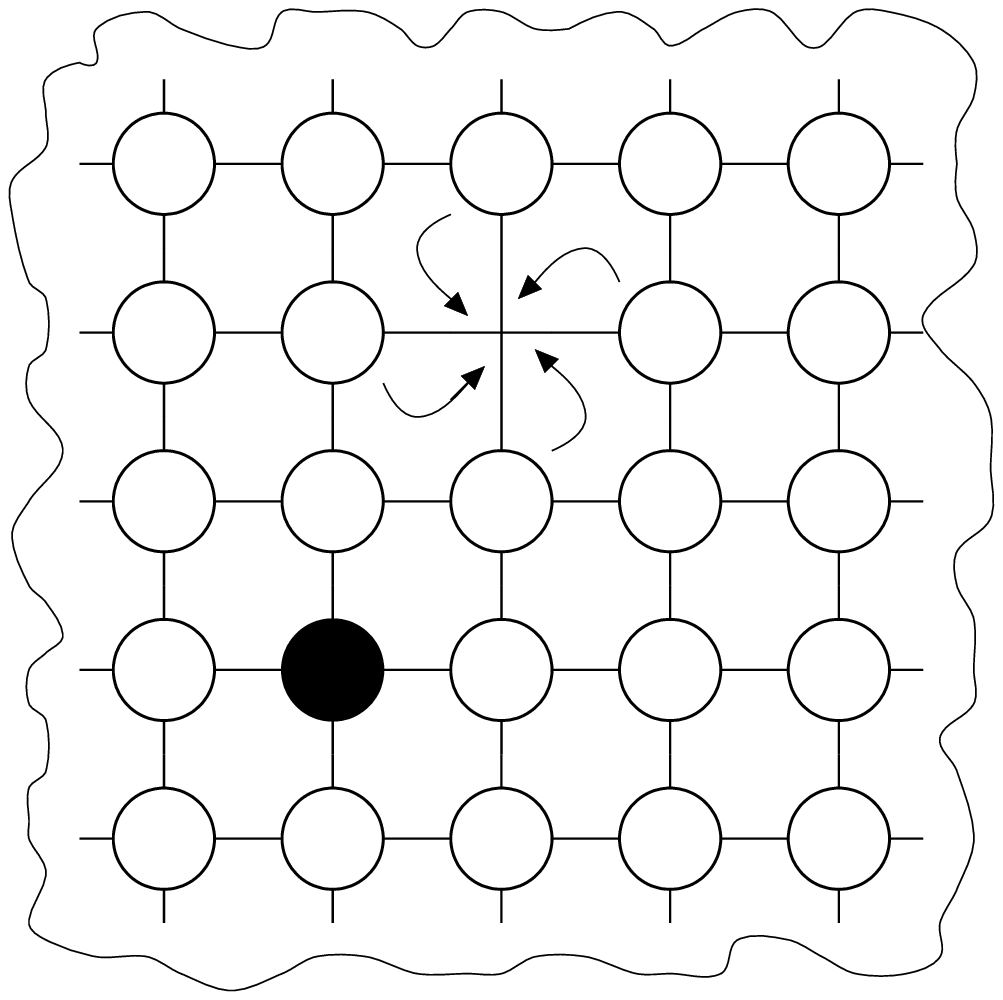}
\vspace*{-26pt}
\fcaption{The infinite sea of particles with
one tagged (the solid circle) particle, and
one empty site.}
\end{figure}\noindent
a single 
``white'' particle (or ball) except for  
two sites: one of them is empty (called 
``hole'' or vacancy) and the other  
is occupied by a ``black'' colored ball,
 or particle, see Fig. 1.

The system evolves according to the 
following rules:\\ 1) There is {\em only}  
an {\em excluded volume constraint} on
 the balls; no other interactions are  
present.\\ 2) The balls can take one 
step (equal to one lattice constant)  
jumps and only onto the empty site. \\ 
3) The probability that the empty  
site will be occupied by one of its 
nearest neighbours is the same for every  
neighbour: $p=1/2d$.\\ 4) The black
 ball is distinguished from the others   
{\em only} by its color. Otherwise 
it obeys exactly the same rules 1)--3) as  
the others.

It is obvious that the hole performs 
random walk of P\'{o}lya type 
\cite{Hughes,Polya1,Polya2}. Due to rule
 2), the black ball performs a  
``passive'' walk, moving only as a 
result of being ``kicked'' by the  
vacancy. One might as well say that 
it is {\em driven} by the Brownian  
vacancy. Our main concern is the 
motion of the colored ball. More precisely  
various properties of the walk 
performed by the black ball will be derived  
rigorously.

In the following we write down the full probability distribution   
for the pair: vacancy and tagged particle.   
We have also found an alternative, and mathematically more 
rigurous derivation\cite{Tezis}   
than the one presented below, however 
the version presented here is much shorter, and perhaps more
intuitive, too.

Without loss of generality let us assume that, initially, the    
black ball is in the origin,   
and the hole is at ${\bf r_0}$    
(${\bf r_0}\neq {\bf 0}$) (${\bf r_0} \in Z^d$).   
After $n$ steps by the hole, the black    
ball is to be found at site $\mbox{\boldmath$\mbox{\boldmath$\rho$}$}$   
and the hole, at site    
$\mbox{\boldmath$\rho$}+{\bf r}$, see Fig. 2.   
($\mbox{\boldmath$\rho$}\;, \;{\bf r} \in Z^d$,   
${\bf r} \neq {\bf 0}$).    
For later convenience, let us denote:  ${\bf R} \equiv   
\mbox{\boldmath$\rho$}+{\bf r}$.   
\begin{figure}[htbp]   
\vspace*{-2cm}   
\hspace*{1.5cm}\epsfxsize = 4 in    
\epsfbox{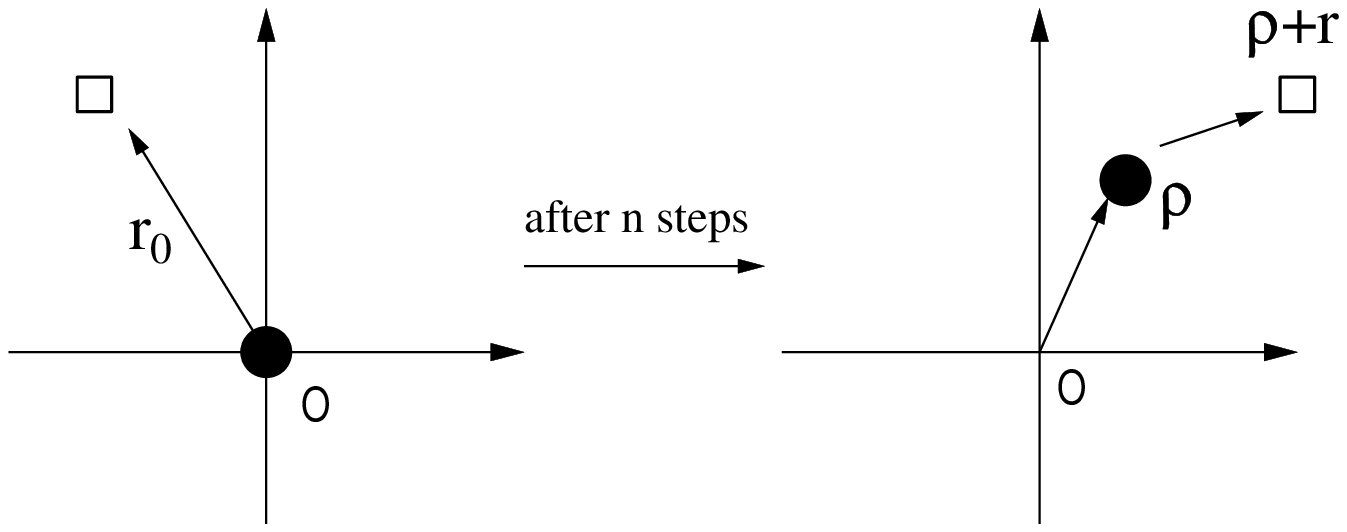}   
\vspace*{-3.5cm}   
\fcaption{Initially the tagged particle (the solid circle)    
is in the origin,   
and the vacancy is at site   
 ${\bf r_0}$ (the empty square).   
After $n$ steps the tagged particle is at site   
$\mbox{\boldmath$\rho$}$ and the random walker at site   
${\bf R}=\mbox{\boldmath$\rho$}+{\bf r}$.}   
\end{figure}   

Let us introduce the following notations:
\begin{itemize}
\item $P_n({\bf s}\!\mid\!{\bf s_0})$ is the
 probability that the vacancy is found at site ${\bf s}$  
after $n$ steps, given it started at ${\bf s_0}$,
$n \geq 1$, and ${\bf s},{\bf s_0} \in Z^d$. For
$n=0$ by convention,
$P_0({\bf s}\!\mid\!{\bf s_0})= 
\delta_{{\bf s},{\bf s_0}}$.

\item $P^{\dagger}_n({\bf s}\!\mid\!{\bf s_0})$ is the probability  
that the vacancy is found at site ${\bf s}$ (${\bf s,s_0}\neq {\bf 0}$)
after $n$ steps {\em in  
the presence of} an irreversible trap at site ${\bf 0}$, given the walk  
commenced at site ${\bf s_0}$. In other words, 
$P^{\dagger}_n({\bf s}\!\mid\!{\bf s_0})$ is just the probability to get to 
${\bf s}$  from ${\bf s_0}$ in $n$  
steps {\em without having visited the origin}. In the literature, the
$P^{\dagger}_n({\bf s}\!\mid\!{\bf s_0})$'s are called `taboo' probabilities,
see Ref. 6. Consistently, 
$P^{\dagger}_{0}({\bf s'}\!\mid\!{\bf s})=   
\delta_{{\bf s'},{\bf s}}$.

\item $\pi^k_n(\mbox{\boldmath$\rho$},   
{\bf r}\!\mid\!{\bf 0},{\bf r_0})$    
is the probability   
that the black ball was driven $k$ steps   
from ${\bf 0}$ to $\mbox{\boldmath$\rho$}$, during which the   
random walker made $n$ steps from ${\bf r_0}$ to   
${\bf r}+\mbox{\boldmath$\rho$}$.
\end{itemize} 

Furthermore, 
$\sum\limits_{{\bf s}}^{({\bf s_0})}$ will mean summing
over all the nearest neighbours of site ${\bf s_0}$,
and in $\sum\limits_{{\bf s}}$ the summation is taken 
over all the lattice sites.

Let us consider   
the referrence frame connected to the tagged particle.   
In this frame the walk performed by the vacancy    
away from the origin is just the regular  P\'olya walk,   
but it can never visit the origin. Instead it ``jumps over''   
the origin, by moving two lattice constants. This jump    
corresponds to the simple exchange of the vacancy and    
tagged particle in the original frame.    
With this observation,    
\begin{eqnarray}   
\pi^k_n(\mbox{\boldmath$\rho$},   
{\bf r}\!\mid\!{\bf 0},{\bf r_0})=p^k   
\sum_{\mbox{{\scriptsize \boldmath$\nu_1$}}}^{({\bf 0})}   
\sum_{\mbox{{\scriptsize \boldmath$\nu_2$}}}^{({\bf 0})}    
 ... \sum_{\mbox{\scriptsize \boldmath$\nu_k$}}^{({\bf 0})}   
\sum_{n_0=1}^{\infty}   
\sum_{n_1=1}^{\infty}   ...     
\sum_{n_k=0}^{\infty}    
\delta_{n,\sum^k_{i=1} n_i}    
\delta_{\mbox{\scriptsize \boldmath$\rho$},\sum^k_{i=1}   
 \mbox{\scriptsize \boldmath$\nu_i$}} && \nonumber \\    
P^{\dagger}_{n_k}({\bf r}\!\mid\!-\mbox{\boldmath$\nu_k$})   
P^{\dagger}_{n_{k-1}-1}(\mbox{\boldmath$\nu_k$}\!\mid\!   
-\mbox{\boldmath$\mbox{\boldmath$\nu_{k-1}$}$})...   
P^{\dagger}_{n_{1}-1}(\mbox{\boldmath$\nu_2$}\!\mid\!   
-\mbox{\boldmath$\mbox{\boldmath$\nu_{1}$}$})
P^{\dagger}_{n_{0}-1}(\mbox{\boldmath$\nu_1$}\!   
\mid\!{\bf r_0})\;,&& \label{pitotal}   \\
k\geq 1,   
\;\;{\bf r_0,r}\neq 0\;. && \nonumber  
\end{eqnarray}   
    
We had to sum over all the
partitions of the time 
$n$ into $k$ subsets, as well
a sum over all the 
possible directions (in number of 
$2d$) from which the 
black ball can be ``hit''.
Obviously if $\mbox{\boldmath$\nu_i$}$ is   
a nearest neighbour of    
the origin, then $-\mbox{\boldmath$\nu_i$}$ is   
another nearest neighbour on    
the other side of the origin,   
straight accross. 

The $k=0$ case has to be treated separately:   
\begin{equation}   
\pi^0_n(\mbox{\boldmath$\rho$},{\bf r}\!\mid\!{\bf 0},{\bf r_0})=   
\delta_{\mbox{\scriptsize \boldmath$\rho$},{\bf 0}}   
P^{\dagger}_{n}({\bf r}\!\mid\!{\bf r_0})\;.  \label{pizero}   
\end{equation}   
By analyzing (\ref{pitotal}), we infer:    
$\pi^k_n(\mbox{\boldmath$\rho$},{\bf r}\!\mid\!{\bf 0},{\bf r_0})=0$,   
if $k > n$,   
as it should be, since the black ball    
cannot be hit more times than the number of steps taken by   
the random walker.

\section{Generating functions}

Usually, it is more convenient to work
with generating functions, instead of the distributions,
themselves. In general, if the generating function for
the distribution $G_n$, i.e., $G(\xi)=\sum_{n=0}^{\infty}\xi^n
G_n$ is known, the $G_n$'s are retrieved by 
the complex integral along the contour 
${\cal C}$ encircling the origin (counterclockwise)::
\begin{equation}  
G_n= \frac{1}{2\pi \mbox{i}} \oint\limits_{\cal C}\frac{d\xi}{\xi^{n+1}}  
G(\xi)\;.  \label{getback}  
\end{equation}  
One can also quote the following relations,
to be used extensively in our treatment: 
\begin{equation}  
\sum_{j=0}^n G_j= \frac{1}{2\pi \mbox{i}}
 \oint\limits_{\cal C}\frac{d\xi}{  
\xi^{n+1}} \frac{1}{1-\xi} G(\xi)\;, \;\;\; \mbox{and}\;\;\;  
\sum_{j=0}^{\infty} G_j= G(1)\;.  \label{partsum}  
\end{equation}
In the case of the P\'olya walk, the generating function
is expressed as \cite{Hughes,Itzykson}:   
\begin{equation}  
P({\bf s}\!\mid\!{\bf s_0};\xi)= \int\limits_{-\pi}^{\pi} \frac{d^dl}{%
(2\pi)^d} \; \frac{e^{\mbox{i}{\bf l}({\bf s}-{\bf s_0})}} {1-\xi\omega({\bf %
l})}\;,  \label{walkint}  
\end{equation}  
where the integral is taken over the first
 Brillouin zone $[-\pi,\pi]^d$,
 and  $\omega$ is the structure function 
 of the walk $\omega({\bf l}) = \frac{1}{d} \sum_{j=1}^{d} cos(l_j)$, $-\pi \leq l_j  
\leq \pi$ 
(${\bf l}=\sum_{j=1}^{d} l_j\; {\bf e_j} $ with $\{{\bf e_j}\}$ as  unit  
vectors in $I\!\!R^d$).

 For the simplicity of writing, let us 
introduce the following notations:
\begin{eqnarray}  
&&t= t(\xi)\equiv P({\bf 0}\!\mid\!{\bf 0};\xi)\;,  \label{t} \\  
&&u= u(\xi)\equiv P(\mbox{\boldmath$\alpha$}\!
\mid\!{\bf 0};\xi)\;,  \label{u} \\  
&&h=h(\xi)\equiv P(\mbox{\boldmath$\alpha$}
 \!\mid\!-\mbox{\boldmath$\alpha$} ;  
\xi)\;,  \label{cross} \\  
&&b=b(\xi)\equiv P(\mbox{\boldmath$\beta$} 
\!\mid\!-\mbox{\boldmath$\alpha$}%
;\xi)\;,\;\;\; \mbox{\boldmath$\beta$} 
\neq \pm \mbox{\boldmath$\alpha$}%
\;,\;\;\; (\mbox{only for}\;\;d\geq 2) 
 \label{simhbt}  
\end{eqnarray}  
where $\mbox{\boldmath$\alpha$}$ and $\mbox{\boldmath$\beta$}$ are nearest  
neighbours of the origin ($\mid\!\mbox{\boldmath$\alpha$}\!\mid=1$, $\mid\!%
\mbox{\boldmath$\beta$}\!\mid=1$ ). $t$ is the generating function for
the return probabilities, $u$ is for the probabilities 
of return to a nearest neighbour, etc.., see Fig. 3.
\begin{figure}[htbp]   
\vspace*{-2cm}   
\hspace*{1.5cm}\epsfxsize = 4 in    
\epsfbox{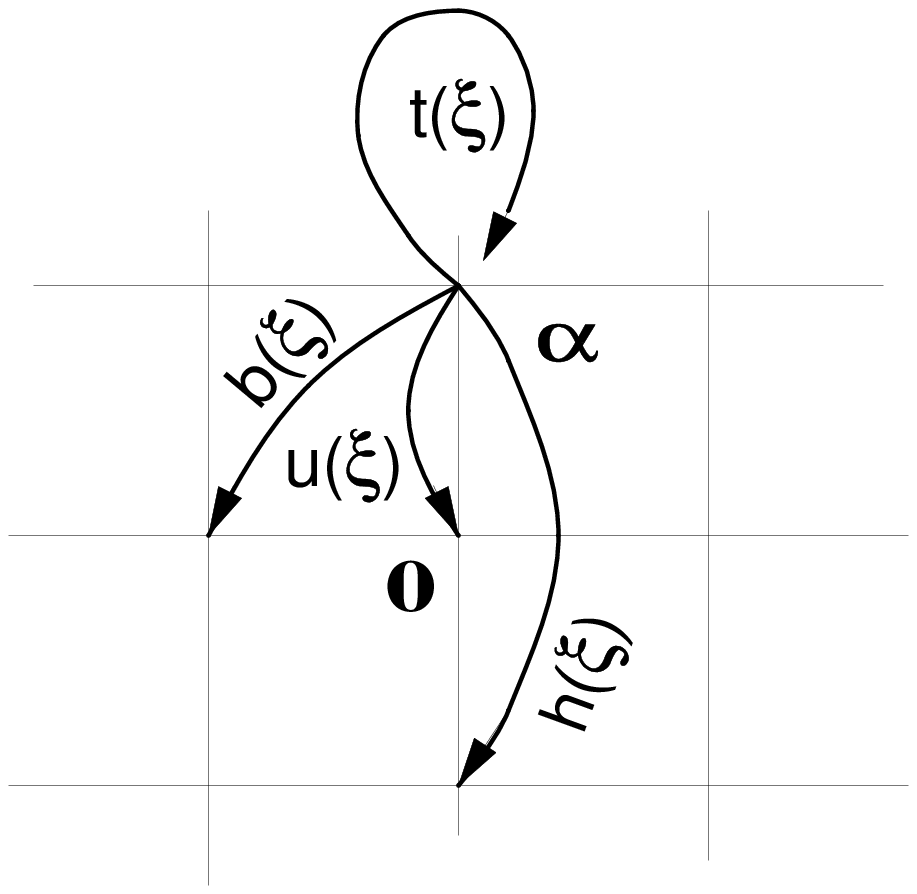}   
\vspace*{-2.5cm}   
\fcaption{The generating functions $t$, $h$, $b$ and $u$.}   
\end{figure}

The connection between the taboo and P\'olya walk probabilities
is most simply given in terms of generating functions\cite{Hughes}:
\begin{equation}  
P^{\dagger}({\bf s}\!\mid\!{\bf s_0};\xi) = P({\bf s}\!\mid\!{\bf s_0};\xi)-   
\frac{1}{t} P({\bf s}\!\mid\!{\bf 0};\xi)\; P({\bf 0}\!\mid\!{\bf s_0}%
;\xi)\;,\;\;\; {\bf s,s_0}\neq {\bf 0}\;.  \label{pdgz}  
\end{equation}
Next, consider the identities\cite{Jones}:   
\begin{eqnarray}  
&&\sum_{{\bf s}} e^{\mbox{i}{\bf l}{\bf s}} = (2\pi)^d \delta({\bf l}%
)\;,\;\;\;\; s\in Z^d\;,\;\;\; {\bf l} \in [-\pi,\pi]^d\;,  \label{20} \\  
&&\sum_{{\scriptsize \mbox{\boldmath$\beta$}}}^{({\bf 0})} e^{\mbox{i} {\bf l%
} \mbox{\scriptsize\boldmath$\beta$}} = \frac{1}{p} \;\omega({\bf l})  
\;,\;\;\;\; {\bf l} \in [-\pi,\pi]^d\;,  \label{21} \\  
&&\int\limits_{-\pi}^{\pi} \frac{d^dl}{(2\pi)^d} e^{\mbox{i}{\bf l}{\bf s}}  
= \delta_{{\bf s},{\bf 0}}\;, \;\;\;\; {\bf s}\in Z^d,\;\;\;{\bf l} \in  
[-\pi,\pi]^d\;.  \label{22}  
\end{eqnarray}
In view of the above, the following useful relations valid for any ${\bf l}  
\in [-\pi,\pi]^d$ and ${\bf s_1},{\bf s_0} \in Z^d$, are easily proved :   
\begin{eqnarray}  
\sum_{{\bf s}} e^{\mbox{i}{\bf l}{\bf s}} P({\bf s}\!\mid\!{\bf s_0};\xi)  
&=& \frac{e^{\mbox{i}{\bf l}{\bf s_0}}} {1-\xi\omega({\bf l})}\;,  
\label{24a} \\  
\sum_{\mbox{\scriptsize \boldmath$\alpha$}}^{({\bf s_1})} e^{\mbox{i}{\bf l} %
\mbox{\scriptsize \boldmath$\alpha$}} P(\mbox{\boldmath$\alpha$}\!\mid\!{\bf %
s_0};\xi) &=& \frac{e^{\mbox{i}{\bf l}{\bf s_1}}}{p\xi} \int\limits_{-\pi}^{%
\pi} \frac{d^d\chi}{(2\pi)^d} \frac{\xi \omega(\mbox{\boldmath$\chi$}-{\bf l}%
)} {1-\xi\omega(\mbox{\boldmath$\chi$})} \; e^{\mbox{i} \mbox{\scriptsize  
\boldmath$\chi$} ({\bf s_1} - {\bf s_0})}\;.  \label{24b}  
\end{eqnarray}  
  
By inserting ${\bf l}={\bf 0}$ in (\ref{24b}) and using (\ref{walkint}) and (%
\ref{22}), is obtained:   
\begin{equation}  
\sum_{\mbox{{\scriptsize   
\boldmath$\alpha$}}}^{({\bf s_1})} P(\mbox{\boldmath$\alpha$} \!\mid\!{\bf %
s_0};\xi) = \frac{1}{p\xi} \Big[ P({\bf s_1}\!\mid\!{\bf s_0} ;\xi)-\delta_{%
{\bf s_1},{\bf s_0}} \Big]\;,  \label{25a}  
\end{equation}  
A further consequence is with ${\bf s_1}={\bf 0}$ ($\mbox{\boldmath$\alpha$}$  
is a nearest neighbour of the origin) and ${\bf s_0}={\bf 0}$ in (\ref{25a}%
). Then, in virtue of notations (\ref{t}-\ref{u}) the left hand side becomes   
$2du$, whereas the right hand side $(t-1)/p\xi$. Thus:   
\begin{equation}  
u=\frac{t-1}{\xi}\;.  \label{ut}  
\end{equation}  
Consider now ${\bf s_1}={\bf 0}$ and ${\bf s_0}\equiv\mbox{\boldmath$\beta$}$  
be a nearest neighbour of the origin. Then from (\ref{25a}):   
\begin{equation}  
\sum_{\mbox{{\scriptsize \boldmath$\alpha$}}}^{({\bf 0})} 
P(\mbox{\boldmath$\alpha$} \!\mid\!\mbox{\boldmath$\beta$};\xi) 
= t+2(d-1)b+h = \frac{u}{p\xi}\;.  \label{ba}  
\end{equation}  
  
Let us now assume the notations: $\sum\limits_{{\bf s}} \!\!\!\mbox{ }
^{\prime}$  will mean summing over all the sites except for ${\bf s}={\bf 0}$%
; $\sum\limits_{{\bf s}}^{({\bf s_1})} \!\!\!\mbox{ }^{\prime}$  will mean  
summing over all the nearest neighbour sites of ${\bf s_1}$ except for ${\bf %
s}={\bf 0}$ (if that is the case).  
  
Similarly to Eqs. (\ref{24a}-\ref{24b}) one can write : for any ${\bf l} \in  
[-\pi,\pi]^d$ and ${\bf s_1},{\bf s_0} \in Z^d$, ${\bf s_0} \neq {\bf 0}$:   
\begin{equation}  
\sum_{{\bf s}} \!\!\!\mbox{ }^{\prime}e^{\mbox{i}{\bf l}{\bf s}} P^{\dagger}  
({\bf s}\!\mid\!{\bf s_0};\xi) = \frac{e^{\mbox{i}{\bf l}{\bf s_0}}}{%
1-\xi\omega({\bf l})} \Bigg[ 1-\frac{e^{\mbox{i}{\bf l}{\bf s_0}}} {t} P(%
{\bf 0}\!\mid\!{\bf s_0};\xi)\Bigg]\;,  \label{31a}  
\end{equation}  
\begin{equation}  
\sum_{\mbox{{\scriptsize \boldmath$\alpha$}}}^{({\bf s_1})} \!\!\!\mbox{ }%
^{\prime}e^{\mbox{i}{\bf l}{\scriptsize \mbox{\boldmath$\alpha$}}}  
P^{\dagger}(\mbox{\boldmath$\alpha$}\!\mid\!{\bf s_0};\xi) = \frac{e^{%
\mbox{i}{\bf l}{\bf s_1} }}{p\xi}\!\! \int\limits_{-\pi}^{\pi} \!\frac{%
d^d\chi}{(2\pi)^d} \frac{\xi \omega(\mbox{\boldmath$\chi$}-{\bf l})} {%
1-\xi\omega(\mbox{\boldmath$\chi$})} e^{\mbox{i}{\scriptsize %
\mbox{\boldmath$\chi$}} ({\bf s_1} - {\bf s_0})}\! \Bigg[ 1-\frac{e^{\mbox{i}  
{\scriptsize \mbox{\boldmath$\chi$}} {\bf s_0}}} {t} P({\bf 0}\!\mid\!{\bf %
s_0};\xi)\Bigg]\;.  \label{31b}  
\end{equation}  
(\ref{31a}) and (\ref{31b}) can easily be proved using (\ref{pdgz}) and (\ref  
{24a}),(\ref{24b}).  
  
A list of useful particular cases:   
\begin{equation}  
\sum_{{\bf s}} \!\!\!\mbox{ }^{\prime}e^{\mbox{i}{\bf l}{\bf s}} P^{\dagger}  
({\bf s}\!\mid\!\mbox{\boldmath$\alpha$};\xi) = \frac{e^{-\mbox{i}{\bf l}   
{\scriptsize \mbox{\boldmath$\alpha$}}}}{1-\xi\omega({\bf l})} \Bigg( 1-%
\frac{u}{t} e^{\mbox{i}{\bf l}{\scriptsize \mbox{\boldmath$\alpha$}}} \Bigg)%
\; ,  \label{32}  
\end{equation}  
\begin{equation}  
\sum_{{\bf s}} \!\!\!\mbox{ }^{\prime}P^{\dagger} ({\bf s}\!\mid\!{\bf s_0}%
;\xi) = \frac{1}{1-\xi} \Bigg[ 1- \frac{P({\bf 0}\!\mid\!{\bf s_0};\xi)} {t} %
\Bigg]\; ,  \label{pker1}  
\end{equation}  
\begin{equation}  
\sum_{{\bf s}} \!\!\!\mbox{ }^{\prime}P^{\dagger} ({\bf s}\!\mid\! %
\mbox{\boldmath$\alpha$};\xi) = \frac{1}{1-\xi} \Bigg( 1-\frac{u}{t} \Bigg)%
\; ,  \label{pker2}  
\end{equation}  
\begin{equation}  
\sum_{{\scriptsize \mbox{\boldmath$\beta$}}}^{({\bf 0})} P^{\dagger}(%
\mbox{\boldmath$\beta$}\!\mid\!{\bf s_0};\xi) = \frac{1}{p \xi} \frac{P({\bf %
0}\!\mid\!{\bf s_0};\xi)}{t}\;,  \label{part3}  
\end{equation}  
and   
\begin{equation}  
\sum_{{\scriptsize \mbox{\boldmath$\beta$}}}^{({\bf 0})} P^{\dagger}(%
\mbox{\boldmath$\beta$}\!\mid\! \mbox{\boldmath$\alpha$};\xi)= \frac{1}{p \xi%
} \frac{u}{t}\;,  \label{part4}  
\end{equation}  
with $\mbox{\boldmath$\alpha$}$ a nearest neighbour of the origin and ${\bf %
s_0} \neq {\bf 0}$.  

Let us now return to our original problem, and
trade the discrete time
variable $n$ for the continuous, complex variable $\xi$,
via:
\begin{equation}   
\Pi^k(\mbox{\boldmath$\rho$},   
{\bf r}\!\mid\!{\bf 0},{\bf r_0};\xi)=   
\sum^{\infty}_{n=0} \xi^n    
\pi^k_n(\mbox{\boldmath$\rho$},{\bf r}\!\mid\!   
{\bf 0},{\bf r_0})\;. \label{pigen1}   
\end{equation}  

From (\ref{pitotal}) and (\ref{pizero})   
 and using the generating    
function for the taboo probabilities    
one finds:   
   
\noindent   
$k\geq 1:$   
\begin{eqnarray}   
\Pi^k(\mbox{\boldmath$\rho$},   
{\bf r}\!\mid\!{\bf 0},{\bf r_0};\xi)=(p\xi)^k   
\sum_{\mbox{{\scriptsize \boldmath$\nu_1$}}}^{({\bf 0})}   
\sum_{\mbox{{\scriptsize \boldmath$\nu_2$}}}^{({\bf 0})}    
 ... \sum_{\mbox{\scriptsize \boldmath$\nu_k$}}^{({\bf 0})}   
\delta_{\mbox{\scriptsize \boldmath$\rho$},   
\sum\limits^k_{i=1} \mbox{\scriptsize \boldmath$\nu_i$}}    
P^{\dagger}({\bf r}\!\mid\!-\mbox{\boldmath$\nu_k$};\xi)   
P^{\dagger}(\mbox{\boldmath$\nu_{k}$}   
\!\mid\!-\mbox{\boldmath$\mbox{\boldmath$\nu_{k-1}$}$};
\xi) ...&& \nonumber \\ 
...P^{\dagger}(\mbox{\boldmath$\nu_{2}$}   
\!\mid\!-\mbox{\boldmath$\nu_1$};\xi)
P^{\dagger}(\mbox{\boldmath$\nu_{1}$}   
\!\mid\!{\bf r_0};\xi)\;,&& \label{Pikx}   
\end{eqnarray}   
$k= 0:$   
\begin{equation}   
\Pi^0(\mbox{\boldmath$\rho$},   
{\bf r}\!\mid\!{\bf 0},{\bf r_0};\xi)=   
\delta_{\mbox{\scriptsize \boldmath$\rho$},0}    
P^{\dagger}({\bf r}\!\mid\!{\bf r_0};\xi)\;.    \label{Pi0x}   
\end{equation}

In Appendix A it is shown that the distribution   
$\pi^k_n(\mbox{\boldmath$\rho$},{\bf r}\!\mid\!   
{\bf 0},{\bf r_0})$ is normalized. (This also serves   
as a check for the correctness of   
(\ref{pitotal}) and (\ref{pizero}).) There we first show   
\begin{equation}   
\sum^{\infty}_{k=0}   
 \sum_{\mbox{{\scriptsize \boldmath$\rho$}}}   
\sum_{{\bf r}} \!\!\!\mbox{ }'      
\delta_{{\bf R},\mbox{\scriptsize \boldmath$\rho$}+{\bf r}}   
 \pi^k_n(\mbox{\boldmath$\rho$},   
 {\bf r}\!\mid\!{\bf 0},{\bf r_0})   
=P_n({\bf R}\!\mid\!{\bf r_0})\;.  \label{norm1}   
\end{equation}   
so that normalization immediately follows 
($\sum_{{\bf R}}P_n({\bf R}\!\mid\!{\bf r_0})=1$).

In many cases, the integral in (\ref{getback}) cannot   
be carried out explicitly. Only asymptotic behavior   
can be given in some limit. Of course, the limit of most interest   
here is $n \rightarrow \infty$. For this purpose,    
we quote an extremely   
useful theorem, the Discrete Tauberian Theorem   
\cite{Hughes,Feller}:   
   
\noindent    
{\bf The Discrete Tauberian Theorem: }    
{\em The relations}   
\begin{equation}   
\sum_{n=0}^{\infty} G_n \xi^n \sim   
(1-\xi)^{-\kappa} L(\frac{1}{1-\xi}) \;\;\;   
\mbox{as}\;\;\; \xi \rightarrow 1^-  \label{taub1}   
\end{equation}   
{\em and}   
\begin{equation}   
G_n \sim \frac{n^{\kappa-1}}   
{\Gamma(\kappa)} L(n)   
\;\;\; \mbox{as}\;\;\; n \rightarrow \infty   
\label{taub2}   
\end{equation}   
{\em are equivalent, provided that }$\kappa>0$;   
$G_n > 0 $ ({\em for large enough} $n$);   
{\em the sequence} $\{G_n\}$    
{\em is monotonic}   
({\em for large enough} $n$);    
{\em and the function} $L$    
{\em is slowly varying in the sense that}   
$\frac{L(\lambda x)}{L(x)} \rightarrow 1   
\;\;\;\mbox{as}\;\;\; x\rightarrow \infty $   
{\em for each fixed, positive} $\lambda$.

\section{Properties of the lattice walk    
performed by  the tagged particle.}

\subsection{The probability that the tagged particle    
suffers $k$ encounters during $n$ steps of the vacancy}   
   
 Let us denote by   
$\hat{\pi}^k_n({\bf r_0})$ the probability that the tagged   
particle has received $k$ hits during $n$ steps of   
the vacancy, given that the vacancy   
commenced at site ${\bf r_0}$   
(${\bf r_0} \neq 0$). One can certainly write:   
\begin{equation}   
\hat{\pi}^k_n({\bf r_0})=\sum_{\mbox{{\scriptsize \boldmath$\rho$}}}    
\sum_{{\bf r}} \!\!\!\mbox{ }'      
\;\pi^k_n(\mbox{\boldmath$\rho$},{\bf r}\!\mid\!{\bf 0},{\bf r_0})\;,    
\label{hat1}   
\end{equation} 
so that the corresponding generating function
$\hat{\Pi}^k({\bf r_0};\xi)=\sum_{n=0}^{\infty}   
\hat{\pi}^k_n({\bf r_0}) \xi^n$ becomes  
\begin{equation}   
\hat{\Pi}^k({\bf r_0};\xi)=   
\sum_{\mbox{{\scriptsize \boldmath$\rho$}}}    
\sum_{{\bf r}} \!\!\!\mbox{ }'\;      
\Pi^k(\mbox{\boldmath$\rho$},{\bf r}\!\mid\!{\bf 0},{\bf r_0};\xi)\;.   
 \label{hat3}   
\end{equation}   
Inserting the expressions   
(\ref{Pikx}),(\ref{Pi0x}) into    
(\ref{hat3}) and then employing   
(\ref{pker2}),(\ref{pker1}) and (\ref{part4}), we can perform the   
sums over ${\bf r}$ and   
 the $\mbox{\boldmath$\nu_j$}$-s. The results are   
   
\noindent   
$k\geq 1:$   
\begin{equation}   
\hat{\Pi}^k({\bf r_0};\xi)=   
\frac{1}{1-\xi}   
\left( 1-   
\frac{u}{t}\right)   
\left(    
\frac{u}{t}\right)^{k-1}   
\frac{P({\bf 0}\!\mid\!{\bf r_0};\xi)}{t}, \label{hatk}   
\end{equation}   
$k= 0:$   
\begin{equation}   
\hat{\Pi}^0({\bf r_0};\xi)=   
\frac{1}{1-\xi}   
\left[1-\frac{1}{t}   
P({\bf 0}\!\mid\!{\bf r_0};\xi)   
\right]. \label{hat0}   
\end{equation}   
Eq. (\ref{hat0}) is the generating function   
for the probability $\hat{\pi}^0_n({\bf r_0})$,   
that the vacancy {\em avoids} the origin during   
the first $n$ steps.   
For $k \geq 1$, the explicit evaluation of the inverse transform  
(\ref{getback}) is possible 
only in one  dimension (see Appendix B for the
expressions of the corresponding   generating functions).     
However, we do not find these results    
particularly illuminating and will not   
present it here. Instead, we analyze    
these probabilities in the $n\rightarrow \infty$   
limit, giving the leading behaviour in $n$ through   
The Discrete Tauberian Theorem.   
   
$\underline{d=1}\;\;$ Using the expressions    
for the generating functions    
in $d=1$ from Appendix B and    
applying the Discrete Tauberian Theorem   
for $k\geq 1$ with $\kappa=1/2$, and   
for $k=0$ with $\kappa=1$, we find   
\begin{equation}   
\hat{\pi}^k_n({\bf r_0})\sim    
\frac{1}{\sqrt{n}} \;\sqrt{\frac{2}{\pi}}   
\;e^{-\sqrt{2} \frac{ k+\mid {\bf r_0} \mid -1}   
{ \sqrt{n}}}    
\;\;\mbox{and for}\;\;   
k = 0:\;\;   
\hat{\pi}^0_n({\bf r_0})\sim1-    
e^{-\sqrt{2} \frac{\mid {\bf r_0} \mid}   
{\sqrt{n}}}    
\label{hatkveg}   
\end{equation}   
   
$\underline{d=2}\;\;$ For simplicity, we
shall treat here the case   
${\bf r_0}=\mbox{\boldmath$\alpha$}$, only, i.e., the vacancy   
starts from a nearest neighbour of the tagged   
particle which is in the origin.    
As we will see, the probability for the tracer   
to suffer $k$ hits scales as $\ln{n}$. Thus we may expect that,   
apart from the initial ``wait'' until the walker   
reaches the tracer for the first time (i.e., if 
${\bf r_0}$ is not necessarily a nearest neighbour),   
the leading behaviour to be independent of which direction   
the vacancy first hits the tagged particle.   
                                             
Thus, we take $P({\bf 0}\!\mid\!{\bf r_0};\xi)=u$.   
Considering the limiting behaviour    
(\ref{limtxi}) for $t$ and    
applying the Tauberian Theorem with   
$\kappa=1$, we arrive at:   
\begin{equation}   
\hat{\pi}^k_n(\mbox{\boldmath$\alpha$})\sim   
\frac{\pi}{\ln{n}}   
\;e^{   -\pi   \frac{k}{\ln{n}}}   
, \;\;\; k \geq 0\;,\;\;\;n\gg1\;. \label{hat2pikveg}   
\end{equation}   
this result shows that the number of hits the tagged particle   
received scales like the $\ln{n}$, with an exponential (or geometric)    
scaling function. Note that, from Section 4.2,    
distances for the Brownian vacancy driven walk do scale as     
the square root of $k$, {\em the number of steps taken by   
the tagged particle}! Of course, the scaling function there   
is not Gaussian.   
   
$\underline{d\geq 3}\;\;$  $t(1)$ is no longer divergent,   
but is a finite number larger than 1.   
One can easily see that for `$n=\infty$':   
\begin{equation}   
\hat{\pi}^k_{\infty}({\bf r_0})=   
\frac{1}{t(1)}   
\left[   
1-\frac{1}{t(1)}   
\right]^{k}   
\frac{P({\bf 0}\!\mid\!{\bf r_0};1)}   
{t(1)-1}.    \label{hat3veg}   
\end{equation}   
If ${\bf r_0}$ is a nearest neighbour of the origin, then   
$\hat{\pi}^k_{\infty}(\mbox{\boldmath$\alpha$})=\frac{1}{t(1)}   
\left[   
1-\frac{1}{t(1)}   
\right]^{k}$.
If $R({\bf 0})$ denotes the probability of return to the origin,
then according to Ref. 6 $R({\bf 0}) = 1-1/t(1)$, thus 
$\hat{\pi}^k_{\infty}(\mbox{\boldmath$\alpha$})
=[1-R({\bf 0})] R({\bf 0})^k$.   
Observe that, independently of   
 dimensionality, the large $k$-behaviour   
is a decaying {\em exponential}.

Another interesting question is:    
`what is the average number of hits   
suffered by the tagged particle during $n$-steps of the walker   
i.e., $\langle k \rangle_n({\bf r_0})$'?    
Naively, one expects that it is simply   
the average number of visits the vacancy pays to the   
origin, when started from site ${\bf r_0}$:    
$\langle I \rangle_n({\bf r_0})$.   
However, on closer examination, we see that   
the tagged particle wanders away from its   
original position (the origin) and {\em towards} the initial position   
of the walker, so that it is not quite at the origin all the time.    
Therefore,    
$\langle k \rangle_n({\bf r_0})$ can differ   
from $\langle I \rangle_n({\bf r_0})$.    
We shall compute the difference $\langle k \rangle_n({\bf r_0})   
-\langle I \rangle_n({\bf r_0})$. This   
is easily given as:   
\begin{equation}   
\langle k \rangle_n({\bf r_0})   
-\langle I \rangle_n({\bf r_0})=   
\frac{1}{2\pi \mbox{i}}    
\oint\limits_{\Gamma}\frac{d\xi}{\xi^{n+1}}   
\;\frac{t-1}{1-(1-\xi)t}\;   
P({\bf 0}\!\mid\!{\bf r_0};\xi)\;. \label{diffhits}   
\end{equation}   
 In one dimension,   
the large $n$-behaviour turns out to be:   
\begin{equation}   
\langle k \rangle_n({\bf r_0})   
-\langle I \rangle_n({\bf r_0})\sim   
\frac{1}{2} \left( 1+\frac{1}{\sqrt{2n}} \right)   
\; \left( 1- \sqrt{\frac{2}{n}}    
\right)^{\mid {\bf r_0}   
\mid +1}\;,    
\end{equation}   
i.e., in the $n \to \infty$ limit    
the tagged particle receives   
$0.5$ {\em more} visits than the origin.   
   
In two dimensions, after    
evaluating the cut in the integral   
(\ref{diffhits}) in leading    
order, the following behaviour   
is found:   
\begin{equation}   
\langle k \rangle_n({\bf r_0})   
-\langle I \rangle_n({\bf r_0})\sim   
\frac{2}{\pi^2}\;\frac{1}{n} [ \psi(n) + c ] \sim    
\frac{2}{\pi^2}\;\frac{\ln{n}}{n}\;,\;\;\;n \gg 1\;,   
\end{equation}   
where $\psi(n)$ is the Euler psi-function and $c$ is the   
Euler constant. After taking the limit $n \to \infty$, one has   
$\langle k \rangle_{\infty}({\bf r_0}) =    
\langle I \rangle_{\infty}({\bf r_0})$.   
   
In $d \ge 3$, there are no singularities   
in the integrand and the difference approaches to zero rapidly   
with $n$.   
   
Observe that  $\langle k \rangle_{n}({\bf r_0}) >   
\langle I \rangle_{n}$, i.e., the tagged   
particle receives {\em more} visits    
than the origin in any dimension.   
This is explained by the following.   
As it will be shown in Section 4.3, the center of the walk performed   
by the tagged particle is always shifted from origin    
towards ${\bf r_0}$ (by an amount    
$\langle\mbox{\boldmath$\rho$}\rangle_{n}   
({\bf r_0})$). Since  ${\bf r_0}$ is the center   
of the walk performed by the vacancy, it means that the   
average position of the tagged particle is {\em always}   
in between the average position of the vacancy and the origin.   
Thus, the tagged particle receives more visits than the origin.

\subsection{The probability that the tagged particle    
 arrives at site $\mbox{\boldmath$\rho$}$    
 on the \protect$n$-th step  of the vacancy}   
   
Let us denote by $\overline{\pi}_n(\mbox{\boldmath$\rho$}   
\!\mid\!{\bf r_0})$ the   
probability that the tagged particle arrives at   
site $\mbox{\boldmath$\rho$}$ on the $n$-th step of the vacancy.   
Clearly, we have   
\begin{equation}   
\overline{\pi}_n(\mbox{\boldmath$\rho$}\!\mid\!   
{\bf r_0})=\sum_{k=0}^{\infty}   
\sum_{{\bf r}} \!\!\!\mbox{ }'    
\pi^k_n(\mbox{\boldmath$\rho$},{\bf r}\!\mid\!{\bf 0},{\bf r_0})\;.    
\label{pt01}   
\end{equation}   
while the corresponding generating function is   
\begin{equation}   
\overline{\Pi}(\mbox{\boldmath$\rho$}\!\mid\!   
{\bf r_0};\xi)=\sum_{k=0}^{\infty}   
\sum_{{\bf r}} \!\!\!\mbox{ }'   
 \Pi^k(\mbox{\boldmath$\rho$},{\bf r}   
 \!\mid\!{\bf 0},{\bf r_0};\xi)\;. \label{pt02}   
\end{equation}   
In this sum, the $k=0$ term has to be treated separately, and it is:   
\begin{equation}   
\delta_{\mbox{\scriptsize \boldmath$\rho$},{\bf 0}} \frac{1}{1-\xi}   
\left[1- \frac{1}{t}P({\bf 0}\!\mid\!{\bf r_0};\xi)   
 \right]\;. \label{pt03}   
\end{equation}   
Let us denote by $\overline{\Pi}_1(\mbox{\boldmath$\rho$}   
\!\mid\!{\bf r_0};\xi)$ the   
rest ($k \geq 1$) of the terms in (\ref{pt02}).    
Using  (\ref{pker2}), the sum over ${\bf r}$   
can readily be performed to give:   
\begin{equation}   
\overline{\Pi}_1(\mbox{\boldmath$\rho$}\!\mid\!{\bf r_0};\xi)=   
\frac{1-u/t}{1-\xi} \sum_{k=1}^{\infty}   
(p\xi)^k   
\sum_{\mbox{{\scriptsize \boldmath$\nu_1$}}}^{({\bf 0})}...    
\sum_{\mbox{{\scriptsize \boldmath$\nu_k$}}}^{({\bf 0})}   
\delta_{{\scriptsize \mbox{\boldmath$\rho$}},   
\sum\limits^k_{j=1} \mbox{{\scriptsize \boldmath$\nu_j$}}}    
P^{\dagger}(\mbox{\boldmath$\nu_{k}$}\!\mid\!   
-\mbox{\boldmath$\mbox{\boldmath$\nu_{k-1}$}$};\xi)...   
P^{\dagger}(\mbox{\boldmath$\nu_{1}$}   
\!\mid\!{\bf r_0};\xi)\;. \label{pt05}   
\end{equation}   
After introducing the function   
\begin{equation}   
\Psi_{{\bf l}}(\mbox{\boldmath$\chi$})=\sum^{\infty}_{k=1} (p \xi)^k   
\sum_{\mbox{{\scriptsize \boldmath$\nu_1$}}}^{({\bf 0})}..    
\sum_{\mbox{{\scriptsize \boldmath$\nu_k$}}}^{({\bf 0})}   
e^{-\mbox{i}({\bf l}- \mbox{{\scriptsize\boldmath$\chi$}})   
\mbox{{\scriptsize \boldmath$\nu_k$}}}    
P^{\dagger}(\mbox{\boldmath$\nu_{k}$}\!\mid\!-   
\mbox{\boldmath$\mbox{\boldmath$\nu_{k-1}$}$};\xi)   
e^{-\mbox{i}{\bf l}\mbox{{\scriptsize \boldmath$\nu_{k-1}$}}}..   
 e^{-\mbox{i}{\bf l}\mbox{{\scriptsize \boldmath$\nu_{1}$}}}   
P^{\dagger}(\mbox{\boldmath$\nu_{1}$}   
\!\mid\!{\bf r_0};\xi) \label{pt07}   
\end{equation}   
and using the integral representation of the Kronecker   
delta, one finds:   
\begin{equation}   
\overline{\Pi}_1(\mbox{\boldmath$\rho$}\!\mid\!{\bf r_0};\xi)=   
\frac{1-u/t}{1-\xi}    
\int\limits_{-\pi}^{\pi} \frac{d^d l}{(2\pi)^d}\;   
e^{-\mbox{i}{\bf l}\mbox{{\scriptsize \boldmath$\rho$}}}   
\Psi_{{\bf l}}({\bf 0})\;.\label{pt06}   
\end{equation}   
   
Let us now evaluate the function $\Psi_{{\bf l}}(\mbox{\boldmath$\chi$})$,   
regarding ${\bf l}$ as a {\em parameter}.   
Define first the following integral operator:   
\begin{equation}   
\hat{\Gamma}_{{\bf l}}^{(\mbox{{\scriptsize \boldmath$\chi$}})}   
(\mbox{\boldmath$\vartheta$}) \;  \equiv   
\int\limits_{-\pi}^{\pi} \frac{d^d \vartheta}{(2\pi)^d}   
\frac{\xi\omega(\mbox{\boldmath$\vartheta$}-{\bf l}+
\mbox{{\boldmath$\chi$}})}   
{1-\xi\omega(\mbox{\boldmath$\vartheta$})} \;  \;\;.    \label{pt09}   
\end{equation} 
The result of the action of 
$\hat{\Gamma}_{{\bf l}}^{(\mbox{{\scriptsize \boldmath$\chi$}})}
(\mbox{\boldmath$\vartheta$})$ on a function of variable
$\mbox{\boldmath$\vartheta$}$  
is a function of a variable $\mbox{{\boldmath$\chi$}}$ 
and parameter ${\bf l}$.
The $k=1$ term in (\ref{pt07}) is   
\begin{equation}   
\hat{\Gamma}_{{\bf l}}^{(\mbox{{\scriptsize    
\boldmath$\chi$}})}(\mbox{\boldmath$\vartheta$})   
 A({\bf r_0};\xi,\mbox{\boldmath$\vartheta$})\;, \label{pt10}   
\end{equation}   
where   
\begin{equation}   
A({\bf r_0};\xi,\mbox{\boldmath$\vartheta$})=    
 e^{-\mbox{i}\mbox{{\scriptsize \boldmath$\vartheta$}}    
 {\bf r_0}}-   
 \frac{1}{t}P({\bf 0}\!\mid\!{\bf r_0};\xi)\;. \label{pt11}   
\end{equation}   
After performing the sum over    
$\mbox{\boldmath$\nu_k$}$  in the rest of   
the terms ($k\geq 2$), and comparing the resultant sums with   
(\ref{pt07}), we find the following integral equation:   
\begin{equation}   
\Psi_{{\bf l}}(\mbox{\boldmath$\chi$})=    
\hat{\Gamma}_{{\bf l}}^{(\mbox{{\scriptsize \boldmath$\chi$}})}   
(\mbox{\boldmath$\vartheta_1$})   
A({\bf r_0};\xi,\mbox{\boldmath$\vartheta_1$})-   
\frac{u}{t}\Psi_{{\bf l}}({\bf 0})   
\hat{\Gamma}_{{\bf l}}^{(\mbox{{\scriptsize \boldmath$\chi$}})}+   
\hat{\Gamma}_{{\bf l}}^{(\mbox{{\scriptsize \boldmath$\chi$}})}   
(\mbox{\boldmath$\vartheta_1$})   
\Psi_{{\bf l}}(\mbox{\boldmath$\vartheta_1$})\;, \label{pt12}   
\end{equation}   
where ${\Gamma}_{{\bf l}}^{(\mbox{{\scriptsize   
 \boldmath$\chi$}})}$ is just    
$\hat{\Gamma}_{{\bf l}}^   
{(\mbox{{\scriptsize \boldmath$\chi$}})}   
(\mbox{\boldmath$\vartheta_1$})$ applied   
on the constant ($=1$) function.   
Substituting the right hand side of (\ref{pt12}) into the    
$\Psi_{{\bf l}}(\mbox{\boldmath$\vartheta_1$})$ on the right   
repeatedly,    
we arrive at:   
\begin{equation}   
\Psi_{{\bf l}}(\mbox{\boldmath$\chi$})=    
\hat{T}_{{\bf l}}^{(\mbox{{\scriptsize    
\boldmath$\chi$}})}(\mbox{\boldmath$\vartheta_1$})   
\hat{\Gamma}_{{\bf l}}^{(\mbox{{\scriptsize    
\boldmath$\vartheta_1$}})}   
(\mbox{\boldmath$\vartheta$})   
A({\bf r_0};\xi,\mbox{\boldmath$\vartheta$})-   
\frac{u}{t}\Psi_{{\bf l}}({\bf 0})\hat{T}_{{\bf l}}^   
{(\mbox{{\scriptsize \boldmath$\chi$}})}   
(\mbox{\boldmath$\vartheta_1$})   
{\Gamma}_{{\bf l}}^{(\mbox{{\scriptsize    
\boldmath$\vartheta_1$}})}\;, \label{pt13}   
\end{equation}   
where    
\begin{eqnarray}   
\hat{T}_{{\bf l}}^{(\mbox{{\scriptsize \boldmath$\chi$}})}   
(\mbox{\boldmath$\vartheta_1$}) \equiv   
\hat{1}^{(\mbox{{\scriptsize \boldmath$\chi$}})}   
(\mbox{\boldmath$\vartheta_1$}) \;    
+\;\hat{\Gamma}_{{\bf l}}^{(\mbox{{\scriptsize    
\boldmath$\chi$}})}(\mbox{\boldmath$\vartheta_1$})&+&   
\hat{\Gamma}_{{\bf l}}^{(\mbox{{\scriptsize    
\boldmath$\chi$}})}(\mbox{\boldmath$\vartheta_2$})   
\hat{\Gamma}_{{\bf l}}^{(\mbox{{\scriptsize \boldmath$\vartheta_2$}})}   
(\mbox{\boldmath$\vartheta_1$}) \nonumber \\
&+&    
\hat{\Gamma}_{{\bf l}}^{(\mbox{{\scriptsize \boldmath$\chi$}})}   
(\mbox{\boldmath$\vartheta_3$})   
\hat{\Gamma}_{{\bf l}}^{(\mbox{{\scriptsize \boldmath$\vartheta_3$}})}   
(\mbox{\boldmath$\vartheta_2$})   
\hat{\Gamma}_{{\bf l}}^{(\mbox{{\scriptsize \boldmath$\vartheta_2$}})}   
(\mbox{\boldmath$\vartheta_1$})   
\;     
...\;, \label{pt14}   
\end{eqnarray}   
with   
\begin{equation}   
\hat{1}^{(\mbox{{\scriptsize    
\boldmath$\chi$}})}   
(\mbox{\boldmath$\vartheta_1$}) \; \;  \equiv   
\int\limits_{-\pi}^{\pi} \frac{d^d    
\vartheta_1}{(2\pi)^d}\;   
\delta(\mbox{\boldmath$\vartheta_1$}-   
\mbox{\boldmath$\chi$})\; \; \;, \label{ptp14}   
\end{equation}   
Choosing $\mbox{\boldmath$\chi$}= {\bf 0}$ in (\ref{pt13}):   
\begin{equation}   
\Psi_{{\bf l}}({\bf 0})=   
\frac{\hat{T}_{{\bf l}}^{({\bf 0})}   
(\mbox{\boldmath$\vartheta_1$})   
\hat{\Gamma}_{{\bf l}}^{(\mbox{{\scriptsize    
\boldmath$\vartheta_1$}})}(\mbox{\boldmath$\vartheta$})   
A({\bf r_0};\xi,\mbox{\boldmath$\vartheta$})   
}{       
1+\frac{u}{t}\hat{T}_{{\bf l}}^{({\bf 0})}   
(\mbox{\boldmath$\vartheta_1$})   
\hat{\Gamma}_{{\bf l}}^{(\mbox{{\scriptsize   
 \boldmath$\vartheta_1$}})}}\;. \label{pt15}   
\end{equation}   
In the following we investigate the operators   
$\hat{\Gamma}$ and $\hat{T}$. By virtue of    
(\ref{21}) and (\ref{pt09}), the action of   
$\hat{\Gamma}$ on a function    
$f:I\!\!R^d \to I\!\!\!\!C$ may be written as:   
\begin{equation}   
\hat{\Gamma}_{{\bf l}}^{(\mbox{{\scriptsize    
\boldmath$\chi$}})}(\mbox{\boldmath$\vartheta$})    
f(\mbox{\boldmath$\vartheta$}) = p\xi   
\sum_{\mbox{{\scriptsize \boldmath$\beta$}}}^{({\bf 0})}   
e^{\mbox{i}\mbox{{\scriptsize \boldmath$\chi$}}   
 \mbox{{\scriptsize \boldmath$\beta$}}}   
\int\limits_{-\pi}^{\pi} \frac{d^d    
\vartheta}{(2\pi)^d}   
\frac{e^{\mbox{i}(\mbox{{\scriptsize \boldmath$\vartheta$}}   
 -{\bf l})\mbox{{\scriptsize \boldmath$\beta$}}}}   
{1-\xi\omega(\mbox{\boldmath$\vartheta$})}\;   
f(\mbox{\boldmath$\vartheta$}) \equiv   
\sum_{\mbox{{\scriptsize \boldmath$\beta$}}}^{({\bf 0})}   
f_{{\bf l} \mbox{{\scriptsize \boldmath$\beta$}}}   
e^{\mbox{i}\mbox{{\scriptsize \boldmath$\chi$}}    
\mbox{{\scriptsize \boldmath$\beta$}}}\;, \label{pt16}    
\end{equation}   
with $f_{{\bf l} \mbox{{\scriptsize \boldmath$\beta$}}}$    
as $2d$, ${\bf l}$-dependent  complex numbers.   
   
Since the $\mbox{\boldmath$\beta$}$'s are lattice   
 sites (elements of $Z^d$), they are of   
the form:   
\begin{eqnarray}   
\mbox{\boldmath$\beta$} = (0,...0,   
&\sigma &,   
0,...,0)\;,\;\;\;   
\sigma \in \{-1,1\}\;, \nonumber \\   
&\stackrel{ \stackrel{     
\uparrow}{  \mbox{ }}}{\mu}&   
\;,\;\;\;\;\;\;\;\;\;\;\;\;\;\;   
\mu \in \{1,2,...,d \} \label{pmu}    
\end{eqnarray}   
($\mu$ labels the axes).   
Then, instead of $\mbox{\boldmath$\beta$}$ one can use the {\em integer}   
$j$ given by the following one-to-one correspondence.   
\begin{equation}   
j=(1-\sigma)\frac{d}{2}+\mu\;, \label{oto}   
\end{equation}   
if $\mbox{\boldmath$\beta$}$ is known. Conversely,    
\begin{eqnarray}   
\sigma = 1 - 2 \left[\frac{j-1}{d} \right]\;,   
\;\;\;\;\mu = j - d\left[\frac{j-1}{d} \right]\;,   \label{mfj}   
\end{eqnarray}   
when $j$ is given.   
An immediate consequence is that when $\sigma=+1$, we have   
$j \in \{ 1,...,d\}$ and when $\sigma = -1$, we have    
$j \in \{ d+1,...,2d \}$.   
   
Let us denote  by $V(I\!\!R^d)$ the set of functions   
$v:I\!\!R^d \to I\!\!\!\!C$ spanned by the basis:   
\begin{equation}   
\tilde{V}(I\!\!R^d)=\left\{ b^1,b^2,...,b^{2d} \right\},\;\;\;   
b^{j}:I\!\!R^d \to I\!\!\!\!C,\;\;\;   
b^{j}(\mbox{\boldmath$\chi$})=   
 e^{\mbox{i}\mbox{{\scriptsize \boldmath$\chi$}}   
  \mbox{{\scriptsize \boldmath$\beta$}} (j)}\;, \label{pt17}   
\end{equation}   
where $\mbox{\boldmath$\beta$} (j)$ is the nearest neighbor of the origin   
corresponding to the integer $j$ via (\ref{mfj})   
and (\ref{pmu}).   
($V(I\!\!R^d)$ forms a linear space with the usual rules of addition   
and multiplication   
of functions over the division ring of the complex numbers.)   
Note from (\ref{pt16}) that the result of the action of    
$\hat{\Gamma}$ on any function    
$f:I\!\!R^d \to I\!\!\!\!C$ is a function from $V(I\!\!R^d)$.   
Furthermore, since the operators $\hat{\Gamma}$ and $\hat{T}$   
are linear operators they admit a representation by $2d \times 2d$    
matrices in this linear space with the basis as in (\ref{pt17}).   
Let ${\bf \Gamma}_{{\bf l}}$ and ${\bf T}_{{\bf l}}$ denote the two matrices.   
Every element $v= \sum\limits_{j=1}^{2d}v_{j} b^{j}$ is represented   
as a $2d \times 1$ columnar matrix $[v]$ in this basis:   
\begin{eqnarray}   
[v] = \left[ \begin{array}{c}   
v_1 \\ . \\ . \\ . \\ v_{2d} \end{array}   
\right]\;.  \label{pt18}   
\end{eqnarray}   
Let us now define the matrix ${\bf P}=\left[ p^{m}_{j} \right]$,   
$j, m \in \{1,...,2d \}$ in this linear space,   
with elements given by:   
\begin{equation}   
p^{m}_{j} = P(\mbox{\boldmath$\beta$}(j)   
 \!\mid\!-\mbox{\boldmath$\alpha$}(m) ;    
\xi)\;. \label{pmat}   
\end{equation}   
Explicitly:   
\begin{eqnarray}   
{\bf P} = \left[ \begin{array}{ccccccccccccc}   
 h & b & . & . & . & b & | & t & b & . & . & . & b \\   
 b & h & . & . & . & b & | & b & t & . & . & . & b \\   
   &   &   & \vdots  &   &   &  | &  &   &   & \vdots  &   &   \\    
 b & b & . & . & . & h & | & b & b & . & . & . & t \\   
 - & - & - & - & - & - & - & - & - & - & - & - & -  \\   
 t & b & . & . & . & b & | & h & b & . & . & . & b  \\   
 b & t & . & . & . & b & | & b & h & . & . & . & b  \\   
   &   &   & \vdots  &   &   & | &   &   &   & \vdots  &   &   \\    
 b & b & . & . & . & t & | & b & b & . & . & . & h \\    
 \end{array}   
\right]\;.  \label{mat01}   
\end{eqnarray}   
Next, let us consider the diagonal    
matrix ${\bf L}_{{\bf l}}=\left[ L^{m}_{{\bf l}j} \right]$,
with matrix elements:   
\begin{equation}   
L^{m}_{{\bf l}j} = e^{-\mbox{i} {\bf l}    
\mbox{{\scriptsize \boldmath$\beta$}}(j)} \delta_{j}^{m}\;, \label{pllmat}   
\end{equation}   
where $\delta_{j}^m$  (Kronecker delta) are   
 the elements of the $2d \times 2d$    
unit matrix ${\bf 1}$.   

The operator $\hat{\Gamma}$ will be represented by
the matrix ${\bf \Gamma}_{{\bf l}}=
\left[ \gamma^{m}_{{\bf l}j} \right]$, with elements:   
\begin{equation}   
\gamma^{m}_{{\bf l}j} =   
p\xi \; \int\limits_{-\pi}^{\pi}    
\frac{d^d \vartheta}{(2\pi)^d}   
\frac{e^{\mbox{i}\mbox{{\scriptsize \boldmath$\vartheta$}}   
 [\mbox{{\scriptsize \boldmath$\alpha$}}(m)+   
 \mbox{{\scriptsize \boldmath$\beta$}}(j)]}}   
{1-\xi\omega(\mbox{\boldmath$\vartheta$})}   
 \;e^{-\mbox{i} {\bf l} \mbox{{\scriptsize 
 \boldmath$\beta$}}(j)}=   
p\xi P(\mbox{\boldmath$\beta$}(j)   
\!\mid\!-\mbox{\boldmath$\alpha$}(m) ; \xi)    
\;e^{-\mbox{i} {\bf l} \mbox{{\scriptsize   
 \boldmath$\beta$}}(j)}\;. \label{pt20}   
\end{equation}   
From the above, we have:   
\begin{equation}   
{\bf \Gamma}_{{\bf l}}= p    
\xi {\bf L}_{{\bf l}} {\bf P}\;. \label{glp}   
\end{equation}   
Then, based on (\ref{pt14}), it is easy to write    
down the representation for the operator   
$\hat{T}$, ${\bf T}_{{\bf l}}=\left[ t^{m}_{{\bf l}j} \right]$:
\begin{equation}   
{\bf T}_{{\bf l}} = {\bf 1}+   
{\bf \Gamma}_{{\bf l}}+{\bf \Gamma}_{{\bf l}}^2+   
{\bf \Gamma}_{{\bf l}}^3+...= \left({\bf 1}-    
{\bf \Gamma}_{{\bf l}} \right)^{-1}\;. \label{pt22}   
\end{equation}   
Examining (\ref{pt15}), it is obvious that we need to calculate   
the action of the operator $\hat{\Gamma}$ on two functions:   
the constant ($=1$) function and    
$e^{-\mbox{i}\mbox{{\scriptsize \boldmath$\vartheta$}} {\bf r_0}}$.   
If $[\iota_{{\bf l}}]$ denotes the vector representing the action   
of $\hat{\Gamma}$ on the function unity and $[\tau_{{\bf l}}]$ represents   
the action of $\hat{\Gamma}$ on    
$e^{-\mbox{i}\mbox{{\scriptsize \boldmath$\vartheta$}} {\bf r_0}}$, we have:   
\begin{eqnarray}   
&& [\iota_{{\bf l}}] = \sqrt{p} \xi u    
{\bf L}_{{\bf l}} [\phi_1]\;, \label{iota1} \\   
&& [\tau_{{\bf l}}] = p \xi    
{\bf L}_{{\bf l}} [Y_0]\;, \label{tauu}   
\end{eqnarray}   
where $[\phi_1]$ is the eigenvector of ${\bf P}$ corresponding   
to the eigenvalue $\lambda_1=u/(p\xi)$ (see Appendix C) and   
$[Y_0]$ is the vector:   
\begin{eqnarray}   
[Y_0] = \left[ \begin{array}{c}   
 P(\mbox{\boldmath$\beta$}(1)\!\mid\! {\bf r_0} ; \xi) \\ \vdots \\    
   P(\mbox{\boldmath$\beta$}(2d)\!\mid\! {\bf r_0} ; \xi) \end{array}   
\right]\;.  \label{Y0}   
\end{eqnarray}   
Note that   
\begin{equation}   
[\phi_1]^T [Y_0] =\frac{1}{\sqrt{p} \xi}    
P({\bf 0}\!\mid\!{\bf r_0};\xi)\;. \label{yoo}   
\end{equation}   
Since 
\begin{eqnarray} 
\hat{T}_{{\bf l}}^{(\mbox{{\scriptsize 
 \boldmath$\chi$}})}(\mbox{\boldmath$\vartheta_1$}) 
{\Gamma}_{{\bf l}}^ 
{(\mbox{{\scriptsize \boldmath$\vartheta_1$}})} & = & 
\sum_{j=1}^{2d} \left( 
{\bf T}_{{\bf l}}\;[\iota_{{\bf l}}]  
\right)_{j} 
 b^{j} 
 (\mbox{{\scriptsize \boldmath$\chi$}})\;, \label{pt25} \\ 
\hat{T}_{{\bf l}}^{(\mbox{{\scriptsize  
\boldmath$\chi$}})}(\mbox{\boldmath$\vartheta_1$}) 
\hat{\Gamma}_{{\bf l}}^{(\mbox{{\scriptsize  
\boldmath$\vartheta_1$}})}(\mbox{\boldmath$\vartheta$}) 
e^{-\mbox{i}\mbox{{\scriptsize \boldmath$\vartheta$}} {\bf r_0}} & = & 
 \sum_{j=1}^{2d} \left( 
{\bf T}_{{\bf l}}\;[\tau_{{\bf l}}]  
\right)_{j}  
b^{j} 
(\mbox{\boldmath$\chi$})\;, \label{pt26} 
\end{eqnarray} 
and $b^{j}(0)=1$,    
we arrive at:   
\begin{equation}   
\Psi_{{\bf l}}({\bf 0})= \frac{       
\sqrt{p} \xi [\phi_1]^T {\bf T}_{{\bf l}} {\bf L}_{{\bf l}}    
[Y_0] - \xi \frac{u}{t}P({\bf 0}\!\mid\!{\bf r_0};\xi)   
[\phi_1]^T {\bf T}_{{\bf l}} {\bf L}_{{\bf l}}[\phi_1]    
}{    1 + \xi \frac{u^2}{t}   
[\phi_1]^T   
{\bf T}_{{\bf l}} {\bf L}_{{\bf l}} [\phi_1] }\;. \label{pt29}   
\end{equation}   
In Appendix C, we show how to diagonalise the matrix  ${\bf P}$    
and write it in the form    
${\bf P}= {\bf C}^{-1} {\bf \Omega}{\bf C}$,   
with ${\bf \Omega}$ a diagonal matrix containing the eigenvalues   
of ${\bf P}$ on the diagonal. The columns of the    
matrix ${\bf C}^{-1}$ are the right eigenvectors of   
${\bf P}$. Therefore   
\begin{eqnarray}   
[\phi_1]^T {\bf C}^{-1} =[1\;0\;...\;0]\;,\;\;\;   
\mbox{and}\;\;\;\;\;{\bf C} [\phi_1]=   
\left[ \begin{array}{c}   
1 \\ 0 \\ \vdots \\ 0 \end{array}   
\right]\;.  \label{cprod}   
\end{eqnarray}   
Thus   
\begin{eqnarray}   
&&[\phi_1]^T{\bf T}_{{\bf l}} {\bf L}_{{\bf l}} [\phi_1]  =    
[\phi_1]^T ({\bf 1} - p\xi {\bf L}_{{\bf l}} {\bf P})^{-1}   
{\bf L}_{{\bf l}}[\phi_1]=    
[\phi_1]^T ({\bf L}_{- {\bf l}}- p\xi {\bf P})^{-1}[\phi_1] =   
\nonumber \\   
&& [\phi_1]^T{\bf C}^{-1} ({\bf C}{\bf L}_{- {\bf l}}   
{\bf C}^{-1}- p\xi{\bf \Omega})^{-1}   
 {\bf C}[\phi_1]   \;, \nonumber   
\end{eqnarray}   
where we have used the fact ${\bf L}_{{\bf l}}^{-1}={\bf L}_{- {\bf l}}$.   
Therefore $[\phi_1]^T{\bf T}_{{\bf l}} {\bf L}_{{\bf l}} [\phi_1]$ is   
nothing but the $m=1$, $j=1$ element of the   
matrix    
$({\bf C}{\bf L}_{- {\bf l}}{\bf C}^{-1}- p\xi{\bf \Omega})^{-1}$.   
   
For $l=0$, we have ${\bf L}_{0}={\bf 1}$, thus the first element   
in the above matrix is:   
\begin{equation}   
[\phi_1]^T{\bf T}_0 {\bf L}_0 [\phi_1]  =    
\frac{1}{1-p\xi \lambda_1}=\frac{1}{1-u}\;.   \label{firsteig}   
\end{equation}   
With the help of (\ref{yoo}),   
the first term in the numerator of (\ref{pt29}) becomes:   
\begin{equation}   
\sqrt{p} \xi [\phi_1]^T {\bf T}_0 {\bf L}_0    
[Y_0] = \frac{1}{1-u}\;   
\frac{1}{t}\;P({\bf 0}\!\mid\!{\bf r_0};\xi)\;, \label{ftn}   
\end{equation}   
and the second:   
\begin{equation}   
\xi \frac{u}{t}\;P({\bf 0}\!\mid\!{\bf r_0};\xi)   
[\phi_1]^T {\bf T}_0 {\bf L}_0[\phi_1]   
= \xi  \frac{u}{t} \;\frac{1}{1-u}\;   
P({\bf 0}\!\mid\!{\bf r_0};\xi)\;.  \label{scnd}   
\end{equation}   
Collecting all the above together, we have:   
\begin{equation}   
\Psi_0(0)= \frac{1}{   1-\frac{u}{t}}\;   
\frac{1}{t}\;P({\bf 0}\!\mid\!{\bf r_0};\xi)\;. \label{el0}   
\end{equation}   
   
One check for the above calculations is to see    
if the distribution $(\ref{pt01})$ is normalized.   
For this purpose, we consider:   
\begin{eqnarray}   
&&\sum_{\mbox{{\scriptsize \boldmath$\rho$}}} \overline{\Pi}   
(\mbox{\boldmath$\rho$}\!\mid\!{\bf r_0};\xi)=   
\frac{1}{1-\xi}\left[1-   
\frac{1}{t}P({\bf 0}\!\mid\!{\bf r_0};\xi) \right] +   
\sum_{\mbox{{\scriptsize \boldmath$\rho$}}} \overline{\Pi}_1   
(\mbox{\boldmath$\rho$}\!\mid\!{\bf r_0};\xi) = \nonumber \\   
&&\frac{1}{1-\xi}\left[1-   
\frac{1}{t}P({\bf 0}\!\mid\!{\bf r_0};\xi) \right]+   
\frac{1-u/t}{1-\xi} \Psi_0(0)=\frac{1}{1-\xi}\;,   
\end{eqnarray}   
showing that normalization is indeed properly satisfied.    
   
Considering the first term (\ref{pt03}) and the expression   
(\ref{el0}), the generating function is:   
\begin{equation}   
\overline{\Pi}(\mbox{\boldmath$\rho$}   
\!\mid\!{\bf r_0};\xi)= \frac{1}{1-\xi}   
\int\limits_{-\pi}^{\pi} \frac{d^d l}{(2\pi)^d}\;   
e^{-\mbox{i}{\bf l}\mbox{{\scriptsize \boldmath$\rho$}}}    
\left\{ 1+\left(1-\frac{u}{t}   
 \right)    
 \Big[ \Psi_{{\bf l}}({\bf 0})-\Psi_{\bf 0}({\bf 0}) \Big]    
 \right\} \;, \label{gfuu}   
\end{equation}   
where $\Psi_{{\bf l}}({\bf 0})$ is given by (\ref{pt29}).

In the following we examine the meaning of formulas   
(\ref{gfuu}) and (\ref{pt29}), by considering them in    
different spatial dimensions.   
   
\noindent    
\underline{$d=1$}.  This case is considered more as a   
check for the validity of our formulas since in one dimensions   
the problem is rather trivial: the random walker   
is either on the left or on the right of the tagged   
particle, and the only way to change sides is to   
exchange positions with the tagged particle. So, if   
initially the tagged particle was in the origin and   
the walker was to the right ($r_0 > 0$),    
the only two sites visited   
by the tagged particle are sites $0$ and $1$.   
Nevertheless, one can ask a nontrivial question:   
`what is the probability to find the tagged particle   
on sites $0$ and $1$ after $n$-steps of the random   
walker, given the walker started from site $r_0$ ?'   
This can be answered in the framework of our theory.   
Without loss of generality, we assume $r_0 > 0$.   
   
The matrix ${\bf P}$ is in the simplest possible form:   
\begin{eqnarray}   
{\bf P} = \left[ \begin{array}{ccc}   
 h & t  \\   
 t & h  \\   
 \end{array}   
\right]\;.  \label{matrd1}   
\end{eqnarray}   
   
The computation of $\Psi_{{\bf l}}({\bf 0})$ is    
rather straightforward using   
(\ref{pt29}) :   
\begin{equation}   
\Psi_{l}(0)=\Psi_{0}(0) \left(    
1- \frac{1-e^{-\mbox{i}l}}{     
1+\frac{u}{t}}\right)\;,   
\label{psi1d}   
\end{equation}   
where $\Psi_0(0)$ is given by (\ref{el0}) and   
\begin{equation}   
P(0\!\mid\!r_0;\xi) =    
u \left( \frac{u}{t} \right)^{r_0 -1 }\;,\;\;\;   
\frac{u}{t} = \frac{1-\sqrt{1-\xi^2}}{\xi}\;. \label{b1d}   
\end{equation}   
See Appendix B for the generating functions.   
The integral over $l$ in (\ref{pt06}) is trivial   
and we finally obtain:   
\begin{equation}   
\overline{\Pi}(\rho   
\!\mid\!r_0;\xi)= \frac{1}{1-\xi}   
\left\{ \left[ 1- \frac{    \left( \frac{u}{t}\right)^{r_0}}   
{   1+\frac{u}{t}} \right] \delta_{\rho,0} +   
\frac{\left(    \frac{u}{t}\right)^{r_0}}   
{   1+\frac{u}{t}}\delta_{\rho,1}    
\right\}\;. \label{pibar1d}   
\end{equation}   
Indeed, as expected, there are only two sites the tagged   
particle may visit! Moreover, using Eqs. (\ref{getback}) and    
(\ref{partsum}), we have $\overline{\pi}_{\infty}(\rho   
\!\mid\!{r_0})=(1/2)(\delta_{\rho,{0}}+   
\delta_{\rho,{1}})$ . In other words, the tagged   
particle spends half of its time on site $1$ and half    
on site $0$. The $n$-dependence is found by simply performing   
the integral with respect to $\xi$.

\noindent    
\underline{$d=2$}. This is the nontrivial case.
 In the theory of recurrent events it is shown under
very general circumstances that the return probability
$R({\bf s_0})$ to a site ${\bf s_0}$ for a lattice walk,   
is simply expressed by: $R({\bf s_0}) = 
1-1/P({\bf s_0}\!\mid\!{\bf s_0};1^{-})$, where
$P({\bf s_0}\!\mid\!{\bf s_0};\xi)$ is the generating function
for the lattice walk (which can be almost anything, not necessarily
Brownian). In Ref. 9 we proved that the same conditions are applicable
to the walk performed by the tagged particle as well, provided we replace
$P_n({\bf s}\!\mid\!{\bf s_0})$ by
$\overline{\pi}_{n}(\mbox{\boldmath$\rho$}
\!\mid\!{\bf r_0})$. Here, $n$ means both `time' and the number
of steps taken by the vacancy, $\mbox{\boldmath$\rho$}$ ``corresponds''
to ${\bf s}$ and, according to our initial conditions, ${\bf s_0}$
is replaced by ${\bf 0}$ while ${\bf r_0}$ serves as a parameter.
Therefore, the walk   
performed by the tagged particle is recurrent if    
$\overline{\Pi}(\mbox{\boldmath$0$}   
\!\mid\!{\bf r_0};1^-)$ is divergent, which is indeed the case, see below.
   
In two dimensions our matrices are four by four, and the   
process of evaluating  $\overline{\Pi}(\mbox{\boldmath$\rho$}   
\!\mid\!{\bf r_0};\xi)$ via (\ref{gfuu}) and (\ref{pt29})   
is not particularly illuminating.   
We report only its final, analytic expression:   
\begin{equation}   
\overline{\Pi}(\mbox{\boldmath$\rho$}   
\!\mid\!{\bf r_0};\xi) =    
\frac{\delta_{\scriptsize \mbox{\boldmath$\rho$},   
{\bf 0}}}{1-\xi}   
\Big[1- \frac{1}{t}P({\bf 0}\!\mid\!{\bf r_0};\xi) \Big]+   
\frac{1}{1-\xi}   
\sum_{j=1}^{4} g_j(\mbox{\boldmath$\rho$};\xi)   
{\cal K}_j({\bf r_0};\xi)\;, \label{fin2d}   
\end{equation}   
where   
\begin{equation}   
g_j(\mbox{\boldmath$\rho$};\xi) =    
\left(1- \frac{u}{t} \right)   
\int\limits_{-\pi}^{\pi} \frac{d^2 l}{(2\pi)^2}   
\;e^{\mbox{i}{\bf l}\mbox{\boldmath$\rho$}}\;   
\frac{\Delta_j({\bf l})}{D({\bf l})}\;,   
\;\;\;j=1,..,4\;, \label{geje}   
\end{equation}   
and   
\begin{eqnarray}   
{\cal K}_1({\bf r_0};\xi) & = &   
\frac{1}{t}P({\bf 0}\!\mid\!{\bf r_0};\xi)\;, \label{kj1} \\   
{\cal K}_2({\bf r_0};\xi) & = &   
\frac{\xi}{4} \Big[ P(\mbox{\boldmath$\beta$}(1)\!\mid\!{\bf r_0};\xi)   
+P(-\mbox{\boldmath$\beta$}(1)\!\mid\!{\bf r_0};\xi) \nonumber \\
&&-    
P(\mbox{\boldmath$\beta$}(2)\!\mid\!{\bf r_0};\xi) -   
P(-\mbox{\boldmath$\beta$}(2)   
\!\mid\!{\bf r_0};\xi) \Big]\;,\label{kj2} \\   
{\cal K}_3({\bf r_0};\xi) & = &   
- \frac{\xi}{2\sqrt{2}}    
\Big[ P(\mbox{\boldmath$\beta$}(2)\!\mid\!{\bf r_0};\xi) -   
P(-\mbox{\boldmath$\beta$}(2)   
\!\mid\!{\bf r_0};\xi) \Big]\;, \label{kj3} \\   
{\cal K}_4({\bf r_0};\xi) & = &   
\frac{\xi}{2\sqrt{2}}    
\Big[ P(\mbox{\boldmath$\beta$}(1)\!\mid\!{\bf r_0};\xi) -   
P(-\mbox{\boldmath$\beta$}(1)   
\!\mid\!{\bf r_0};\xi) \Big]\;. \label{kj4}    
\end{eqnarray}   
The quantities $\Delta_j({\bf l})$ are proportional to the   
elements (up to a common factor) of the first row of the matrix    
$({\bf C}{\bf L}_{- {\bf l}}{\bf C}^{-1}   
- p\xi{\bf \Omega})^{-1}$ and they are   
all expressed in terms of $V = V(\xi) =    
\frac{1}{\xi}[\frac{4}{\pi}{\bf E}(\xi)-1-(1-\xi^2)\frac{2}{\pi}   
{\bf K}(\xi)]$ and $W = W(\xi) =    
\frac{1}{\xi}[\frac{2}{\pi}{\bf E}(\xi)-1]$   
 which are    
the second and third eigenvalues of matrix ${\bf P}$, $\lambda_2$   
and $\lambda_3$ up to a factor $p\xi=\xi/4$,    
Here, ${\bf K}(\xi)$ and ${\bf E}(\xi)$   
 being the complete elliptic integrals   
of first and second kind, respectively (see Appendices C and B).   
\begin{eqnarray}   
\Delta_1({\bf l}) &=& -(1+V W)W + \Big[(1+V W) + (V+W)W \Big]   
\frac{1}{2}\Big[cos(l_1)+cos(l_2)\Big] \nonumber \\   
&&\hspace*{5.5cm} - (V+W)cos(l_1) cos(l_2)\;, \label{delt1} \\   
\Delta_2({\bf l}) &=& (1-W^2)   
\frac{1}{2}\Big[cos(l_1)+cos(l_2)\Big] \;,\label{delt2} \\    
\Delta_3({\bf l}) &=& \frac{\mbox{i}}{\sqrt{2}} sin(l_2)   
\Big[(1+V W) - (V+W)cos(l_1)\Big] \;, \label{delt3} \\   
\Delta_4({\bf l}) &=& -\frac{\mbox{i}}{\sqrt{2}} sin(l_1)   
\Big[(1+V W) - (V+W)cos(l_2)\Big] \;, \label{delt4}    
\end{eqnarray}   
and   
\begin{eqnarray}   
D({\bf l})&=&(1+V W)\left(1+W\frac{u}{t}\right)    
\Big\{ 1- (A+B)\frac{1}{2}\Big[cos(l_1)+cos(l_2)\Big]\nonumber \\
&&\hspace*{5.5cm} +ABcos(l_1) cos(l_2) \Big\}\;, \label{deel}   
\end{eqnarray}   
with   
\begin{equation}   
A= \frac{   W+\frac{u}{t}}   
{   1+W\frac{u}{t}}\;,\;\;\;   
B= \frac{V+W}{1+VW}\;. \label{AB}   
\end{equation}   
Formulas (\ref{fin2d}-\ref{AB}) give the {\em exact} expression   
of the generating function $\overline{\Pi}(\mbox{\boldmath$\rho$}   
\!\mid\!{\bf r_0};\xi)$.    
From (\ref{fin2d}) we see that the    
functions $g_j$ carry the information   
on the spatial component of the walk    
performed by the tagged particle   
while the functions ${\cal K}_j$ describe the    
dependence on the initial condition of the vacancy.   
Rewriting the cosines   
and sines in formulas (\ref{delt1}-   
\ref{delt4}) in terms of exponentials,   
it becomes clear that one  needs in general    
to evaluate the $\xi \to 1^-$   
behaviour of integrals of type:   
\begin{equation}   
\Gamma({\bf s};\xi) = \int\limits_{-\pi}^{\pi}    
\frac{d^2 l}{(2\pi)^2}   
\;\frac{e^{\mbox{i}{\bf l}{\bf s}}}{D({\bf l})}\;.\label{Gam}   
\end{equation}   
By further introducing the vectors   
\begin{eqnarray}    
[\Gamma (\mbox{\boldmath$\rho$};\xi)]_{n.} \equiv   
\left[ \begin{array}{c}   
 \Gamma(\mbox{\boldmath$\rho$}+\mbox{\boldmath$\beta$}(1); \xi) \\    
 \Gamma(\mbox{\boldmath$\rho$}+\mbox{\boldmath$\beta$}(2); \xi) \\    
 \Gamma(\mbox{\boldmath$\rho$}-\mbox{\boldmath$\beta$}(1); \xi) \\    
 \Gamma(\mbox{\boldmath$\rho$}-\mbox{\boldmath$\beta$}(2); \xi)    
  \end{array}   
\right],\;\;    
[\Gamma (\mbox{\boldmath$\rho$};\xi)]_{n.n.} \equiv   
\left[ \begin{array}{c}   
 \Gamma(\mbox{\boldmath$\rho$}+\mbox{\boldmath$\beta$}(1)+   
 \mbox{\boldmath$\beta$}(2); \xi) \\    
 \Gamma(\mbox{\boldmath$\rho$}-\mbox{\boldmath$\beta$}(1)+   
 \mbox{\boldmath$\beta$}(2); \xi) \\    
 \Gamma(\mbox{\boldmath$\rho$}-\mbox{\boldmath$\beta$}(1)-   
 \mbox{\boldmath$\beta$}(2); \xi) \\    
 \Gamma(\mbox{\boldmath$\rho$}+\mbox{\boldmath$\beta$}(1)-   
 \mbox{\boldmath$\beta$}(2); \xi)    
  \end{array}   
\right]     
\end{eqnarray}   
characterizing the nearest   
neighbour and next to nearest neighbour terms corresponding to   
 site $\mbox{\boldmath$\rho$}$, the   
 $g_j$ functions are simply expressed as:   
\begin{eqnarray}   
g_1(\mbox{\boldmath$\rho$};\xi) =    
&-&(1+VW)W\left(1-\frac{u}{t}\right)    
\Gamma (\mbox{\boldmath$\rho$};\xi)+ \nonumber \\   
&&\frac{1}{2}\Big[(1+V W) + (V+W)W \Big]   
\left(1-\frac{u}{t}\right) [\phi_1]^T    
[\Gamma (\mbox{\boldmath$\rho$};\xi)]_{n.}   
- \nonumber \\   
&&\frac{1}{2}(V+W)\left(1-\frac{u}{t}\right)[\phi_1]^T    
[\Gamma (\mbox{\boldmath$\rho$};\xi)]_{n.n.}\;.  \label{ujg1}\\   
&& \nonumber \\   
g_2(\mbox{\boldmath$\rho$};\xi) =    
&-&\frac{1}{2}(1-W^2)\left(1-\frac{u}{t}\right)   
[\phi_2]^T [\Gamma (\mbox{\boldmath$\rho$};\xi)]_{n.}\;, \label{ujg2} \\   
&&\nonumber \\   
g_3(\mbox{\boldmath$\rho$};\xi) =   
&-&\frac{1}{2}(1+V W)\left(1-\frac{u}{t}\right)[\phi_3]^T   
[\Gamma (\mbox{\boldmath$\rho$};\xi)]_{n.}+ \nonumber \\   
&+&\frac{1}{4}(V+W)\left(1-\frac{u}{t}\right)   
\Big( [\phi_3]^T+[\phi_4]^T\Big)   
[\Gamma (\mbox{\boldmath$\rho$};\xi)]_{n.n.}\;, \label{ujg3} \\   
&&\nonumber \\   
g_4(\mbox{\boldmath$\rho$};\xi) =   
& &\frac{1}{2}(1+V W)\left(1-\frac{u}{t}\right)[\phi_4]^T   
[\Gamma (\mbox{\boldmath$\rho$};\xi)]_{n.}+ \nonumber \\   
&+&\frac{1}{4}(V+W)\left(1-\frac{u}{t}\right)   
\Big( [\phi_4]^T-[\phi_3]^T\Big)   
[\Gamma (\mbox{\boldmath$\rho$};\xi)]_{n.n.}\;. \label{ujg4}   
\end{eqnarray}   
The $\xi \to 1^-$ behaviour is determined by the singularities   
of $D({\bf l})$. Analyzing the functions $A$ and $B$    
one can rigurously show that there is only   
one zero of $D({\bf l})$ in the first Brillouin zone, namely   
${\bf l}=0$.   
   
For ${\bf s}=0$, (\ref{Gam}) can be evaluated {\em exactly}:   
\begin{equation}   
\Gamma({\bf 0};\xi) = \frac{2}{\pi} \; \frac{1}{(1+VW)   
(   1+W\frac{u}{t})}   
\;\frac{1}{1-AB} \;{\bf K}\left(\frac{A-B}{1-AB}\right) \;.\label{Gam0}   
\end{equation}   
For ${\bf s} \neq 0$ the exact evaluation of the integral is cumbersome   
(it can be given recursively and in terms of Legendre functions   
of second kind) and it is not our aim here. Instead we analyze the   
$\xi \to 1^-$ behaviour. In this limit $u/t \to 1$,    
$V \to \overline{V}=   
\frac{4}{\pi}-1 \sim 0.273$, $W \to \overline{W}=   
\frac{2}{\pi}-1 \sim -0.363$,   
 $A \to \overline{A}=1$, $B \to \overline{B} \sim -0.100$, 
 see also Appendices B and C.   
 The limiting behaviours for both ${\bf s}=0$ and ${\bf s} \neq 0$   
 cases are computed to be:   
 \begin{equation}   
\Gamma({\bf 0};\xi) \sim \frac{\pi^2}{4} \;    
\frac{1}{(\pi-2)(\pi-1)}   
\;{\bf K}\left(1-\frac{2}{\pi-2}(1-u/t)\right) \;,   
\;\;\;\xi \to 1^-\;,\label{Gamlim0}   
\end{equation}   
 and   
 \begin{equation}   
\Gamma({\bf s};\xi) \sim \frac{\pi^2}{4} \;    
\frac{1}{(\pi-2)(\pi-1)}   
\;K_0\left(2\sqrt{\pi-1} \mid\!{\bf s}\!\mid    
\sqrt{1-u/t}\right) \;,\;\;\;\xi \to 1^-\;,   
\;\;\;{\bf s} \neq 0.\label{Gamlims}   
\end{equation}   
with $K_0$ being a modified Bessel function.   
Let us now keep $\mid\!{\bf s}\!\mid$ constant.   
 Since : $K_0(z) \sim -\ln{z}$ as $z \to 0$ and    
${\bf K}(z) \sim \frac{1}{2} \ln{(\frac{8}{1-z})}$ as $z \to 1^-$ it follows   
that both  $\Gamma$'s {\em diverge}   
and  have the same leading term, namely:   
 \begin{equation}   
\Gamma({\bf s};\xi) \sim \frac{\pi^2}{8} \;    
\frac{1}{(\pi-2)(\pi-1)}   
\; \ln{\left( \frac{1}{   1-\frac{u}{t}}\right)}\;,   
\;\;\; \xi \to 1^-\;.\label{limbeh}   
\end{equation}   
Taking into account that, in (\ref{ujg2}-\ref{ujg4}),   
$[\phi_j]^T [\Gamma (\mbox{\boldmath$\rho$};\xi)]_{n.}$   
and $[\phi_j]^T [\Gamma (\mbox{\boldmath$\rho$};\xi)]_{n.n.}$   
for $j \geq 2$ involve {\em differences} of $\Gamma$'s,    
we see that the nearest neighbour terms and the next nearest neighbour   
terms (and therefore $g_2$, $g_3$, $g_4$) are negligible compared to    
the terms in  $g_1$. Thus, to leading order, we have:   
\begin{equation}   
g_1(\mbox{\boldmath$\rho$};\xi) =    
\left(1-\frac{1}{\pi}\right) \left(1-\frac{u}{t}\right)   
\ln{\left( \frac{1}{    1-\frac{u}{t}}\right)}+   
{\cal O}\left(1-\frac{u}{t} \right)\;,   
\;\;\; \xi \to 1^-\;,\label{limg1}   
\end{equation}   
\begin{equation}   
g_j(\mbox{\boldmath$\rho$};\xi) =    
0+{\cal O}\left(1-\frac{u}{t} \right)\;,   
\;\;\; \xi \to 1^-\;,\;\;\;j=2,3,4\;.\label{limgj}   
\end{equation}   
One can see that the leading  one is $g_1(\mbox{\boldmath$\rho$};\xi)$.   
Although the generating functions    
$P({\bf s}\!\mid\!{\bf r_0};1^-)$ are divergent in two dimensions,   
the quantities ${\cal K}_j({\bf r_0};1^-)$ are all finite numbers.   
This is the consequence of formulas (\ref{kj1}-\ref{kj4}) and the   
fact that when $\xi \to 1^-$ the probability generating functions   
for the P\'olya walk {\em all} diverge in the same way, more specifically   
$\lim\limits_{\xi \to 1^-}[t(\xi)-P({\bf s}\!\mid\!{\bf r_0};\xi)]$   
is a {\em finite} number,   
for any ${\bf s}$ and ${\bf r_0}$ (see for e.g. Ref. 6, p.145).    
Thus ${\cal K}_1({\bf r_0};1^-)=1$. Using this, it also can be    
proved that the $k=0$  term (\ref{pt03})   
 is negligible, compared to the $g_1 {\cal K}_1$ term, in this limit.   
   
Collecting the above, our conclusion is: when    
$\mbox{\boldmath$\rho$}$ is kept finite,    
the generating function $\overline{\Pi}(\mbox{\boldmath$\rho$}   
\!\mid\!{\bf r_0};\xi)$ is,   
to leading order,   
\begin{equation}    
\overline{\Pi}(\mbox{\boldmath$\rho$}   
\!\mid\!{\bf r_0};\xi) \sim \left(1-\frac{1}{\pi} \right)   
\frac{1}{1-\xi} \left(1-\frac{u}{t} \right)   
\ln{\left( \frac{1}{    1-\frac{u}{t}}\right)}\;,   
\;\;\;\xi \to 1^-\;, \;\;\;   
 \mbox{\boldmath$\rho$}\;\;\mbox{finite}\;.  \label{apprx}    
\end{equation}   
Using the Discrete Tauberian Theorem and the fact that:   
\begin{eqnarray}   
1-\frac{u}{t} \sim \frac{\pi}   
{\ln{\left(   \frac{1}{1-\xi}\right)}}\;,\;\;\;   
\mbox{as}\;\;   
0<1-\xi \ll 1\;, \nonumber   
\end{eqnarray}   
we find:   
\begin{equation}   
\overline{\pi}_n(\mbox{\boldmath$\rho$}   
\!\mid\!{\bf r_0}) \sim    
(\pi -1)\frac{\ln{(\ln{n})}}{\ln{n}}\;,\;\;\; n \to \infty\;,   
\;\;\;\mbox{\boldmath$\rho$}   
\;\;\;\mbox{finite}\;. \label{limitbehn}   
\end{equation}   
From (\ref{apprx}),    
$\overline{\Pi}({\bf 0}\!\mid\!{\bf r_0};1^-)$ is clearly   
divergent, proving our assertion that the walk in two dimensions   
is reccurent. The tagged particle is guaranteed to visit any site   
in the $n \to \infty$ limit.

{\em Scaling.} It is well known that, for the pure random (P\'olya)
walk, distances scale as the square root of time, while the scaling
function is, apart from a prefactor, Gaussian. However,
for the Brownian vacancy driven walk, the scaling properties    
are quite different.    
When $\mid \!\mbox{\boldmath$\rho$}\!\mid$ and $n$
are {\em both} large (keep ${\bf r_0}$ constant), 
the only surviving term is again $g_1 {\cal K}_1$.
Since ${\cal K}_1({\bf r_0};1^-)=1$, we infer   
\begin{eqnarray}    
\overline{\Pi}(\mbox{\boldmath$\rho$}\!\mid\!{\bf r_0};\xi)   
 \sim    
\frac{2}{\pi}(\pi -1) \;\frac{1}{1-\xi}\;   
\left(1-\frac{u}{t} \right) \;   
K_0\left(2\sqrt{\pi-1} \mid\!\mbox{\boldmath$\rho$}\!\mid    
\sqrt{1-u/t}\right) \;, && \nonumber \\
\xi \to 1^-\;,   
\;\;\mid \!\mbox{\boldmath$\rho$}\!\mid \gg 1\;,&& \label{scalingg}   
\end{eqnarray}   
or 
\begin{equation}
\overline{\pi}_n(\mbox{\boldmath$\rho$}
\!\mid\!{\bf r_0}) \sim 
2 (\pi -1) \frac{1}{\ln{n}}
\;K_0\left(2\sqrt{\pi(\pi-1)} \frac{\mid\!\mbox{\boldmath$\rho$}\!\mid} 
{\sqrt{\ln{n}}}\right)\;,\;\;\;n \to \infty\;,
\;\;\mid \!\mbox{\boldmath$\rho$}\!\mid \gg 1\;. \label{scaling2d} 
\end{equation}
This is the scaling function in the corresponding limit.
As we see, {\em instead of being Gaussian}, it is 
the modified Bessel function $K_0$. 
From its argument, it follows that distances scale 
like square root of {\em the log of time}.

In the following Section we calculate the first and second 
moments for the displacement of the tagged particle. 
 However, instead of using the distribution 
 $\overline{\pi}_n(\mbox{\boldmath$\rho$}
\!\mid\!{\bf r_0})$ derived in this Section, we will
use a slightly different, but easier, way to arrive at the desired results.

\subsection{Average displacement and standard deviation   
of the tagged particle. }

Let us define the first and second    
moments associated with the tagged particle's displacement   
$\mbox{\boldmath$\rho$}$ up to time $n$:   
\begin{equation}   
\langle\mbox{\boldmath$\rho$}\rangle_n({\bf r_0})=   
\sum_{\mbox{{\scriptsize    
\boldmath$\rho$}}} \mbox{\boldmath$\rho$}   
\overline{\pi}_n(\mbox{\boldmath$\rho$}   
\!\mid\!{\bf r_0})=   
\sum_{\mbox{{\scriptsize    
\boldmath$\rho$}}} \mbox{\boldmath$\rho$} \sum_{k=0}^{\infty}   
\sum_{{\bf r}} \!\!\!\mbox{ }'\;   
\pi^k_n(\mbox{\boldmath$\rho$},   
{\bf r}\!\mid\!{\bf 0},{\bf r_0})\;, \label{1stm}   
\end{equation}   
and   
\begin{equation}   
\langle\mbox{\boldmath $\rho$}^2\rangle_n({\bf r_0})=   
\sum_{\mbox{{\scriptsize    
\boldmath $\rho$}}} \mbox{\boldmath $\rho$}^2    
\overline{\pi}_n(\mbox{\boldmath $\rho$}   
\!\mid\!{\bf r_0})=   
\sum_{\mbox{{\scriptsize    
\boldmath $\rho$}}} \mbox{\boldmath $\rho$}^2    
\sum_{k=0}^{\infty}   
\sum_{{\bf r}} \!\!\!\mbox{ }'\;   
\pi^k_n(\mbox{\boldmath $\rho$},   
{\bf r}\!\mid\!{\bf 0},{\bf r_0})\;, \label{2ndm}   
\end{equation}   
respectively.
The first of these is, of course, 
the average displacement for the particle:
$\langle
\mbox{\boldmath$\rho$}\rangle_n({\bf r_0})$.
The second is
used in finding the ``spread'', i.e., the standard deviation:
\begin{equation}   
{\cal D}_n({\bf r_0})\equiv \sqrt{   
\langle\mbox{\boldmath$\rho$}^2\rangle_n({\bf r_0})-   
[\langle\mbox{\boldmath$\rho$}   
\rangle_n({\bf r_0})]^2}\;. \label{spread}   
\end{equation}   
Instead of computing directly the sums above   
we turn to generating functions   
\begin{equation}   
{\bf A}({\bf r_0};\xi) \equiv    
\sum_{n=0}^{\infty}   
\langle\mbox{\boldmath$\rho$}\rangle_n({\bf r_0}) \xi^n=    
\sum_{k=0}^{\infty}    
\sum_{\mbox{{\scriptsize \boldmath$\rho$}}} \mbox{\boldmath$\rho$}    
\sum_{{\bf r}} \!\!\!\mbox{ }'\;    
\Pi^k(\mbox{\boldmath$\rho$},   
{\bf r}\!\mid\!{\bf 0},{\bf r_0})\;, \label{1stmg}   
\end{equation}   
and   
\begin{equation}   
S({\bf r_0};\xi)\equiv    
\sum_{n=0}^{\infty}   
\langle\mbox{\boldmath$\rho$}^2\rangle_n({\bf r_0}) \xi^n =   
\sum_{k=0}^{\infty}    
\sum_{\mbox{{\scriptsize    
\boldmath$\rho$}}} \mbox{\boldmath$\rho$}^2    
\sum_{{\bf r}} \!\!\!\mbox{ }'\;    
\Pi^k(\mbox{\boldmath$\rho$},{\bf r}   
\!\mid\!{\bf 0},{\bf r_0})\;. \label{2ndmg}   
\end{equation}   
By peforming the sums over ${\bf r}$ and    
$\mbox{\boldmath$\rho$}$   
we arrive at the final expressions for   
${\bf A}({\bf r_0};\xi) $ and $S({\bf r_0};\xi)$    
(see Appendix D):   
\begin{equation}   
{\bf A}({\bf r_0};\xi) =\frac{1}{1-\xi}\;    
\frac{1}{1-q}\;    
{\bf M}({\bf r_0};\xi)\;, \label{Axi}   
\end{equation}   
\begin{equation}    
S({\bf r_0};\xi)=   
\frac{1}{1-\xi}\;    
\frac{1}{1- u/t}\;    
\frac{1+q}   
{1-q}\;    
\frac{P({\bf 0}\!\mid\!{\bf r_0};\xi)}{t}\;, \label{Sxi}   
\end{equation}   
where   
\begin{equation}   
{\bf M}({\bf r_0};\xi) = p\xi   
\sum_{\mbox{{\scriptsize \boldmath$\nu$}}}^{({\bf 0})}
\mbox{\boldmath$\nu$}   
P(\mbox{\boldmath$\nu$}\!\mid\!{\bf r_0};\xi)\;, \label{Mxi}   
\end{equation}   
and   
\begin{equation}   
q=p\xi(h-t)\;. \label{qxi}   
\end{equation}   
Note, for ${\bf r_0}=\mbox{\boldmath$\alpha$}$   
in (\ref{Mxi}) we have:   
\begin{equation}   
{\bf M}(\mbox{\boldmath$\alpha$};\xi)=-q   
 \mbox{\boldmath$\alpha$}. \label{Ma}   
\end{equation}   
The corresponding moments will then be expressed by:   
\begin{equation}   
\langle\mbox{\boldmath$\rho$}\rangle_n({\bf r_0})=   
\frac{1}{2\pi \mbox{i}}    
\oint\limits_{\Gamma}\frac{d\xi}{\xi^{n+1}}   
\frac{1}{1-\xi}   
\;  \frac{{\bf M}({\bf r_0};\xi) }   
{1-q}, \label{1stmexpr}   
\end{equation}   
and   
\begin{equation}   
\langle\mbox{\boldmath$\rho$}^2\rangle_n({\bf r_0})=   
\frac{1}{2\pi \mbox{i}}    
\oint\limits_{\Gamma}\frac{d\xi}{\xi^{n+1}}   
\frac{1}{1-\xi}\;\frac{1}{1- u/t}\;\frac{1+q}{1-q}\;    
\frac{P({\bf 0}\!\mid\!{\bf r_0};\xi)}{t}. \label{2ndmexpr}   
\end{equation}
There is a simple expression\cite{Zia} for ${\bf M}({\bf r_0};\xi)$
showing that it is in fact a vector pointing along ${\bf r_0}$:
\begin{equation}
{\bf M}({\bf r_0};\xi) = {\bf r_0} \frac{1}{\xi}
\int\limits_{0}^{\xi} P({\bf 0}\!\mid\!
{\bf r_0};\zeta) d\zeta\;.  \label{mr0}
\end{equation}
This means, that the average of the tagged walk is always shifted
{\em towards} the initial position of the vacancy.

In the following we extract    
the $n \rightarrow \infty$ behaviour for the different cases.   
   
\noindent   
$\underline{d=1}\;\;\;$   Without loss of generality, assume   
$r_0 \geq 1$. Then, from Appendix B:   
\begin{equation}    
{\bf M}({\bf r_0};\xi)    
=\left( \frac{1-\sqrt{1-\xi^2}}{\xi}   
\right)^{{r_0}} \mbox{{\bf e}}_1\;,    
\;\;\; \mbox{and}\;\;\; q=-    
\frac{1-\sqrt{1-\xi^2}}{\xi}\;. \label{1stm1d2}   
\end{equation}   
Therefore, for the average displacement, we find:   
\begin{equation}   
\langle\mbox{\boldmath$\rho$}\rangle_n({\bf r_0})=   
\frac{\mbox{{\bf e}}_1}{2\pi \mbox{i}}    
\oint\limits_{\Gamma}\frac{d\xi}{\xi^{n}}   
\frac{1}{1-\xi}   
\;  \frac{1}{    
1+\xi+\sqrt{1-\xi^2}}   
\;  \left(   
\frac{1}{1+\sqrt{1-\xi^2}}   
\right)^{r_0-1}. \label{1stm1d4}   
\end{equation}   
Using (\ref{partsum}), we find immediately    
$\langle\rho \rangle_{\infty}({ r_0})=1/2$,   
which is certainly correct because   
half of the time the tagged particle is   
at the origin, and half of the time is   
on site 1. The convergence    
to $1/2$, i.e., the $n$-behaviour in   
leading order is given by:   
\begin{equation}   
\langle\rho\rangle_n({ r_0})\sim   
\frac{1}{2}\left(    
e^{   -\sqrt{2} \frac{r_0-1}   
{\sqrt{n}}} \right)\;\mbox{{\bf e}}_1\;.  \label{1stm1d6}   
\end{equation}   
Observe that, as pointed out below Eq. (\ref{mr0})
 on average, the tagged particle   
suffers a {\em finite displacement toward}   
the initial position of the vacancy.   
One can consider this as an effective   
{\em attraction} between the tagged particle and   
the random walker.   
This is property  valid in any   
dimension.   
   
The second order moment turns out to be:   
$\langle\rho^2\rangle_n({ r_0})=   
\langle\rho\rangle_n({ r_0})$. Thus, the spread is    
${\cal D}_{\infty}({r_0})=1/2$,    
as expected.

$\underline{d=2}\;\;\;$ To obtain explicit expressions 
for the general case is non-trivial.   
For example, if the tagged particle and the vacancy are initially    
very far apart, we expect that the particle will    
experience the first hit to come from all directions with    
equal probability. Then its average displacement will vanish.   
So, we choose to present only the case where the vacancy-tracer   
is initially a nearest neighbour pair, i.e.,   
${\bf r_0}=\mbox{\boldmath$\alpha$}$, where    
$\mbox{\boldmath$\alpha$}$ is a nearest neighbour   
of the origin.   
   
The average displacement is then given by:   
\begin{equation}   
\langle\mbox{\boldmath$\rho$}\rangle_n(\mbox{\boldmath$\alpha$})=   
\frac{1}{2\pi \mbox{i}}    
\oint\limits_{\Gamma}\frac{d\xi}{\xi^{n+1}}   
\frac{1}{1-\xi}   
\;  \frac{(-q)}   
{1-q} \mbox{\boldmath$\alpha$}, \label{1stm2d1}   
\end{equation}   
and in the limit $n\rightarrow \infty$ (see Appendix B for the   
expression of $q$):   
\begin{equation}   
\langle\mbox{\boldmath$\rho$}\rangle_{\infty}(\mbox{\boldmath$\alpha$})=   
\frac{1}{2}\frac{\pi-2}{\pi-1}\mbox{\boldmath$\alpha$}=   
0.266529 \mbox{\boldmath$\alpha$}.   
\end{equation}   
Applying The Discrete Tauberian Theorem,   
the second moment is found:   
\begin{equation}   
\langle\mbox{\boldmath$\rho$}^2   
\rangle_n(\mbox{\boldmath$\alpha$})\sim   
\frac{1}{\pi(\pi-1)} \ln{n}   
\frac{1}{    1-    
\frac{\ln{n}}{n\pi}} \;,   
\;\;\; n \gg 1. \label{2ndm2d2}   
\end{equation}   
this leads to the spread:   
\begin{equation}   
{\cal D}_n(\mbox{\boldmath$\alpha$})\sim   
\left[ \frac{1}{\pi(\pi-1)} \ln{n}-   
\frac{1}{4}\left(   
\frac{\pi-2}{\pi-1}\right)^2   
\right]^{1/2}\sim   
\frac{1}{\sqrt{\pi(\pi-1)}} \sqrt{\ln{n}}   
\;,\;\; n \gg 1. \label{spreadd2}   
\end{equation}   
i.e. it diverges as $\sqrt{\ln{n}}$. This is exactly the result   
we anticipated from the scaling given by (\ref{scaling2d}).   
   
$\underline{d \geq 3}\;\;\;$  Since there 
is no divergence in this case,   
the first and second moments are finite   
even for $n=\infty$:   
\begin{equation}   
\langle\mbox{\boldmath$\rho$}    
\rangle_{\infty}({\bf r_0})=   
\frac{M({\bf r_0};1)}{1-q(1)}\;,\;\;\mbox{and}\;\;   
\langle\mbox{\boldmath$\rho$}^2   
\rangle_{\infty}({\bf r_0})=   
\frac{1+q(1)}{1-q(1)}   
P({\bf 0}\!\mid\!{\bf r_0};1). \label{ujvmi}  
\end{equation}   
The spread,   
\begin{equation}   
{\cal D}_{\infty}({\bf r_0})\sim   
\left\{    
\frac{1+q(1)}{1-q(1)}   
P({\bf 0}\!\mid\!{\bf r_0};1)-   
\left[   
\frac{M({\bf r_0};1)}{1-q(1)}   
\right]^2   
\right\}^{1/2}\; \label{spreadd3}   
\end{equation}   
is also just a finite number.   
   
If the vacancy again started from a nearest neighbour   
$\mbox{\boldmath$\alpha$}$   
of the tagged particle,   
then, we have, via (\ref{Ma}):   
\begin{equation}   
\langle\mbox{\boldmath$\rho$}    
\rangle_{\infty}(\mbox{\boldmath$\alpha$})=   
-\frac{q(1)}{1-q(1)} 
\mbox{\boldmath$\alpha$}\;,\;\;\mbox{and}\;\;       
{\cal D}_{\infty}(\mbox{\boldmath$\alpha$}) =   
\frac{1}{1-q(1)}\;   
\sqrt{\;\Big[ 1- q(1)^2 \Big] \;t(1) -1}\;. \label{neighspreadd3}   
\end{equation}   
In particular, in three dimensions    
$t(1)=1.516386$  and $q(1)=-0.209..$ (see Appendix B), so that   
\begin{equation}   
\langle\mbox{\boldmath$\rho$}    
\rangle_{\infty}(\mbox{\boldmath$\alpha$})=   
0.1728..\;\mbox{\boldmath$\alpha$}\;,\;\;\mbox{and}\;\;       
{\cal D}_{\infty}(\mbox{\boldmath$\alpha$}) =   
0.5549...\;. \label{numericspreadd3}   
\end{equation}

\section{Conclusions and summary}

We investigated the walk performed by   
the tagged particle in the presence of
 a single vacancy in a d-dimensional,   
infinite square lattice. The only 
``active'' object is the vacancy, while   
the tagged particle performs a ``passive''
 walk, being shifted only when it   
exchanged positions with the Brownian 
vacancy. We constructed the complete   
probability distribution, which is 
quite complex, incorporating three   
spatial variables (${\bf r_0}$, 
$\mbox{\boldmath$\rho$}$ and ${\bf r}$) and   
two temporal ones ($n$ and $k$). However, 
physically relevant informations   
can be extracted from different 
projections of the full distribution. An   
example is $\overline{\pi }_n(\mbox{\boldmath
$\rho$}\!\mid \!{\bf r_0})$,   
the probability distribution for the 
tagged particle to be at site $%
\mbox{\boldmath$\rho$}$ on the $n$-th 
step of the vacancy. Another is $\hat{%
\pi}_n^k({\bf r_0})$, the probability 
that it has made $k$ steps during this   
time. Closed form expressions for the
 corresponding generating functions, in   
terms of the generating function of 
the Brownian walker alone, were   
obtained. Exploiting the powerful 
Discrete Tauberian Theorem, we find the   
limiting behaviours of the distributions
 themselves, as $n\rightarrow \infty    
$. We showed that the Brownian vacancy 
driven walk is reccurent in two   
dimensions only. Other interesting 
results associated with a tagged particle   
in $d=2$ are \\  
{\em (i)} the number of steps it takes 
scales as $\ln n$ (as $n\to \infty $)\\   
{\em (ii)} the limiting displacement-
distribution is not a Gaussian but rather   
the modified Bessel function $K_0$\\   
{\em (iii)}  its standard deviation scales 
as $\sqrt{\ln n}$, in contrast to the   
usual $\sqrt{n}$ scaling Brownian walks   
   
By computing the first moment for the 
displacement-distribution of the   
tagged particle, we find that there 
is an effective attraction towards the   
vacancy which shifts the average position
 of the tagged particle from its   
initial point. In $d$=1, 2, and 3, 
this shift is, respectively, 0.5,   
0.266529.., 0.1728.. in units of the 
lattice spacing. For higher dimensions   
this is given by the first expression
in (\ref{ujvmi}). We have seen, that an   
immediate consequence of this 
attraction is the tagged particle being   
visited more frequently than its 
original site. However the difference   
vanishes as $n\rightarrow \infty $, 
for $d\geq 2$.   
   
In $d\geq 3$, the standard deviation 
of the displacement-distribution {\em %
does not} diverge as $n\to \infty $. 
Instead, it converges to a small   
number, e.g., 0.5549 in $d=3$.   

\section{Acknowledgements}  
  
Illuminating discussions with R. K. P. Zia, 
B. Schmittmann, G. Korniss, C. Laberge, and 
S. Sandow are gratefully acknowledged.
This research is supported in part by the US  
National Science Foundation through the
Division of Materials Research and  
the Hungarian Science Foundation under 
grant numbers OTKA F17166 and T17493.

\appendix

\noindent
{\bfit Normalization}\\

When proving (\ref{norm1}) 
it is better to switch to generating
functions. (\ref{norm1}) is equivalent to:
\begin{equation}
\sum^{\infty}_{k=0} \sum_{\mbox{{\scriptsize 
\boldmath$\rho$}}}
\sum_{{\bf r}} \!\!\!\mbox{ }'   
\delta_{{\bf R},{\mbox{\scriptsize 
\boldmath$\rho$}}+{\bf r}} 
\Pi^k(\mbox{\boldmath$\rho$},{\bf r}
\!\mid\!{\bf 0},{\bf r_0};\xi)
=P({\bf R}\!\mid\!{\bf r_0};\xi)\;. \label{norm3}
\end{equation}
Separate the $k=0$ term in (\ref{norm3}):
\begin{eqnarray}
{\cal T}_0\equiv\sum_{\mbox{
{\scriptsize \boldmath$\rho$}}}
\sum_{{\bf r}} \!\!\!\mbox{ }'   
\delta_{{\bf R},{\scriptsize 
\mbox{\boldmath$\rho$}}+{\bf r}}  
\Pi^0(\mbox{\boldmath$\rho$},
{\bf r}\!\mid\!{\bf 0},{\bf r_0};\xi)
\stackrel{(\ref{Pi0x})}{=}
P^{\dagger}({\bf R}\!\mid\!{\bf r_0};\xi) 
\stackrel{(\ref{pdgz})}{=}&& \nonumber \\
P({\bf R}\!\mid\!{\bf r_0};\xi)-\frac{1}{t}
P({\bf R}\!\mid\!{\bf 0};\xi)
P({\bf 0}\!\mid\!{\bf r_0};\xi)\;.&& 
\label{norm4}
\end{eqnarray}    
As for the rest ($\mbox{\boldmath$\rho$}
={\bf R}-{\bf r}$):
\begin{eqnarray}
{\cal T}\equiv\sum^{\infty}_{k=1}
 \sum_{\mbox{{\scriptsize \boldmath$\rho$}}}
\sum_{{\bf r}} \!\!\!\mbox{ }'   
\delta_{{\bf R},{\scriptsize 
\mbox{\boldmath$\rho$}}+{\bf r}} 
\Pi^k(\mbox{\boldmath$\rho$},
{\bf r}\!\mid\!{\bf 0},{\bf r_0};\xi)=
\int\limits_{-\pi}^{\pi} \frac{d^dl}{(2\pi)^d}
e^{\mbox{{\scriptsize i}}{\bf l}{\bf R}}
\sum^{\infty}_{k=1} (p\xi)^k
\sum_{\mbox{{\scriptsize 
\boldmath$\nu_1$}}}^{({\bf 0})}
 ... \sum_{\mbox{{\scriptsize 
 \boldmath$\nu_k$}}}^{({\bf 0})}&& \nonumber \\
\left[ 
\sum_{{\bf r}} \!\!\!\mbox{ }'   
e^{-\mbox{{\scriptsize i}}{\bf l}{\bf r}}
P^{\dagger}({\bf r}\!\mid\!
-\mbox{\boldmath$\nu_k$};\xi)\right]
e^{-\mbox{{\scriptsize i}}{\bf l}
\mbox{{\scriptsize \boldmath$\nu_k$}}}
P^{\dagger}(\mbox{\boldmath$\nu_{k}$}\!\mid\!
-\mbox{\boldmath$\mbox{\boldmath$\nu_{k-1}$}$};\xi)
... 
e^{-\mbox{{\scriptsize i}}{\bf l}
\mbox{{\scriptsize \boldmath$\nu_{1}$}}}
P^{\dagger}(\mbox{\boldmath$\nu_{1}$}
\!\mid\!{\bf r_0};\xi)\;.&& \label{norm5}
\end{eqnarray}
Using 
Eq. (\ref{32})
with $\mbox{\boldmath$\alpha$}=-
\mbox{\boldmath$\nu_k$}$,
the term in square bracket is computed,
and: 
\begin{eqnarray}
{\cal T}&=&
\int\limits_{-\pi}^{\pi} \frac{d^dl}{(2\pi)^d}
\frac{e^{\mbox{{\scriptsize i}}{\bf l}{\bf R}}}
{1-\xi\omega({\bf l})}
\sum^{\infty}_{k=1} (p\xi)^k
\sum_{\mbox{{\scriptsize \boldmath$\nu_1$}}}^{({\bf 0})}
\sum_{\mbox{{\scriptsize \boldmath$\nu_2$}}}^{({\bf 0})} 
 ... \sum_{\mbox{{\scriptsize \boldmath$\nu_k$}}}^{({\bf 0})}
 \left[ 1-\frac{u}{t}
e^{-\mbox{{\scriptsize i}}{\bf l}
\mbox{{\scriptsize \boldmath$\nu_k$}}}
\right] \nonumber \\
&&P^{\dagger}(\mbox{\boldmath$\nu_{k}$}\!\mid\!
-\mbox{\boldmath$\mbox{\boldmath$\nu_{k-1}$}$};\xi)
...
e^{-\mbox{{\scriptsize i}}{\bf l}
\mbox{{\scriptsize \boldmath$\nu_{2}$}}}
P^{\dagger}(\mbox{\boldmath$\nu_{2}$}
\!\mid\!-\mbox{\boldmath$\nu_1$};\xi) 
e^{-\mbox{{\scriptsize i}}{\bf l}
\mbox{{\scriptsize \boldmath$\nu_{1}$}}}
P^{\dagger}(\mbox{\boldmath$\nu_{1}$}
\!\mid\!{\bf r_0};\xi)\;. \label{norm7}
\end{eqnarray}
According to the square bracket
in (\ref{norm7}), ${\cal T}$ splits in two terms
(${\cal T}=\theta_1+\theta_2$).
Using notation (\ref{pt07}), the second term becomes: 
\begin{equation}
\theta_2\equiv
-\frac{u}{t}
\int\limits_{-\pi}^{\pi} \frac{d^dl}{(2\pi)^d}
\frac{e^{\mbox{{\scriptsize   i}}{\bf l}{\bf R}}}
{1-\xi\omega({\bf l})}
\Psi_{\bf l}({\bf 0})\;, \label{theta1}
\end{equation}
while the first:
\begin{eqnarray}
\theta_1 \equiv
\int\limits_{-\pi}^{\pi} \frac{d^dl}{(2\pi)^d}
\frac{e^{\mbox{{\scriptsize i}}{\bf l}{\bf R}}}
{1-\xi\omega({\bf l})}
\sum^{\infty}_{k=1} p^k \xi^k
\sum_{\mbox{{\scriptsize \boldmath$\nu_1$}}}^{({\bf 0})}
\sum_{\mbox{{\scriptsize \boldmath$\nu_2$}}}^{({\bf 0})} ... 
\left[\sum_{\mbox{{\scriptsize \boldmath$\nu_k$}}}^{({\bf 0})}
P^{\dagger}(\mbox{\boldmath$\nu_{k}$}\!\mid\!-
\mbox{\boldmath$\mbox{
\boldmath$\nu_{k-1}$}$};\xi)\right] && \nonumber\\
e^{\mbox{{\scriptsize  i}}{\bf l}
\mbox{\boldmath$\mbox{{\scriptsize \boldmath$\nu_{k-1}$}}$}}
P^{\dagger}(\mbox{\boldmath$
\mbox{\boldmath$\nu_{k-1}$}$}\!\mid\!-
\mbox{\boldmath$\nu_{k-2}$};\xi)
...
e^{-\mbox{{\scriptsize i}}{\bf l}
\mbox{{\scriptsize \boldmath$\nu_{1}$}}}
P^{\dagger}(\mbox{\boldmath$\nu_{1}$}
\!\mid\!{\bf r_0};\xi)\;.&& \label{norm9}
\end{eqnarray}
We again need to separate the $k=1$ term in
$\theta_1$:
\begin{equation}
\theta_1=\int\limits_{-\pi}^{\pi} 
\frac{d^dl}{(2\pi)^d}
\frac{e^{\mbox{{\scriptsize i}}{\bf l}{\bf R}}}
{1-\xi\omega({\bf l})}
p \xi
\left\{ \sum_{\mbox{{\scriptsize
 \boldmath$\nu_1$}}}^{({\bf 0})}
P^{\dagger}(\mbox{\boldmath$\nu_{1}$}\!\mid\!{\bf r_0};\xi)
\right\} + 
(k \geq 2\;\;\mbox{{  terms}})\\\;. \label{tsep}
\end{equation}
The term in square brackets in
(\ref{norm9}) is computed via
(\ref{part4}), while the term in
curly brackets in (\ref{tsep}) is
expressed by (\ref{part3}).
Combining these, we finally get
for the first term:
\begin{eqnarray}
\theta_1 &=&
\frac{1}{t}
P({\bf R}\!\mid\!{\bf 0};
\xi)P({\bf 0}\!\mid\!{\bf r_0};\xi)+
\frac{u}{t}
\int\limits_{-\pi}^{\pi} 
\frac{d^dl}{(2\pi)^d}
\frac{e^{\mbox{{\scriptsize i}}{\bf l}{\bf R}}}
{1-\xi\omega({\bf l})}
\sum^{\infty}_{k=2} (p\xi)^{k-1} 
\sum_{\mbox{{\scriptsize \boldmath$\nu_1$}}}^{({\bf 0})}
\sum_{\mbox{{\scriptsize \boldmath$\nu_2$}}}^{({\bf 0})} ... 
\sum_{\mbox{{\scriptsize \boldmath$\nu_{k-1}$}}}^
{({\bf 0})} \nonumber \\
&&e^{\mbox{{\scriptsize i}}{\bf l}
\mbox{{\scriptsize \boldmath$\nu_{k-1}$}}}
P^{\dagger}(\mbox{\boldmath$\mbox{\boldmath$\nu_{k-1}$}$}
\!\mid\!-\mbox{\boldmath$\nu_{k-2}$};\xi)
...
e^{-\mbox{{\scriptsize i}}{\bf l}
\mbox{{\scriptsize \boldmath$\nu_{1}$}}}
P^{\dagger}(\mbox{\boldmath$\nu_{1}$}\!\mid\!{\bf r_0};\xi) \nonumber \\
&=&
\frac{1}{t}
P({\bf R}\!\mid\!{\bf 0};\xi)P({\bf 0}
\!\mid\!{\bf r_0};\xi)+
\frac{u}{t}
\int\limits_{-\pi}^{\pi} \frac{d^dl}{(2\pi)^d}
\frac{e^{\mbox{{\scriptsize i}}{\bf l}{\bf R}}}
{1-\xi\omega({\bf l})}
\Psi_{\bf l}({\bf 0})\;. \label{norm12}
\end{eqnarray}
Adding the first ($\theta_1$) 
and the second term ($\theta_2$), is obtained:
\begin{equation}
{\cal T}=
\frac{1}{t}
P({\bf R}\!\mid\!{\bf 0};\xi)
P({\bf 0}\!\mid\!{\bf r_0};\xi)\;. \label{qed1}
\end{equation}
(the terms containing the $\Psi$-s cancel).
Add (\ref{qed1}) to (\ref{norm4}) and after 
cancellation, (\ref{norm3}) is gained.
Therefore normalization (\ref{norm3}) holds.

\appendix

\noindent
{\bfit Generating functions}\\

For dimension one, the explicit expression
for the $P(0\!\mid\!R;\xi)$ generating function
is well known \cite{Hughes}:
\begin{equation}
P(0\!\mid\!R;\xi)=
t
\left( \frac{1-\sqrt{1-\xi^2}}{\xi}
\right)^{\mid R \mid}\;,  \label{pxi1d}
\end{equation}
with:
\begin{equation}
t=P(0\!\mid\!0;\xi)=
(1-\xi^2)^{-\frac{1}{2}}\;. \label{pxi01d}
\end{equation}
Thus, according to (\ref{u}) (or 
(\ref{ut}) ), and (\ref{cross}):
\begin{equation}
u=t \frac{1-\sqrt{1-\xi^2}}{\xi}
\;,\;\;\;h=\frac{u^2}{t}\;. \label{196}
\end{equation}

In two dimensions \cite{Hughes,Montroll}:
\begin{equation}
t=P({\bf 0}\!\mid\!{\bf 0};\xi)=\frac{2}{\pi}
{\bf K}(\xi)\;, \label{hatt0}
\end{equation}
where ${\bf K}(\xi)$ is the 
complete elliptic integral
of first kind.
Then, according to (\ref{ut}):
\begin{equation}
u=\frac{1}{\xi}\left[ \frac{2}
{\pi}{\bf K}(\xi)-1\right]\;.
\label{u2d}
\end{equation}
Quoting Ref. 15, the probabilities 
$P({\bf 0}\!\mid\!{\bf R};\xi)$ with 
${\bf R}={\bf R}(m,m)$, $m\in Z$ (${\bf R}(m_1,m_2)$ 
is a lattice vector with components $m_1$ and $m_2$)
can be expressed in terms of Legendre functions
of second kind:
\begin{equation}
P({\bf 0}\!\mid\!{\bf R}(m,m);\xi)=\frac{2}{\pi \xi}
Q_{m-1/2}\left(
\frac{1-\xi^2/2}{\xi^2/2}
\right)\;. \label{mm}
\end{equation}
Note, that for the P\'olya walk
$P({\bf 0}\!\mid\!{\bf R}(m_1,m_2); \xi) =
P({\bf 0}\!\mid\!{\bf R}( m_1,-m_2); \xi) =
P({\bf 0}\!\mid\!{\bf R}(-m_1,m_2); \xi) =
P({\bf 0}\!\mid\!{\bf R}(-m_1,-m_2); \xi)$ for any 
$m_1,m_2 \in Z$.

For $m=0$ we regain (\ref{hatt0}).
For $m=1$:
\begin{equation}
b\equiv P({\bf 0}\!\mid\!R(1,1);\xi)
=\frac{4}{\pi \xi^2}\left[
\left(1-\frac{\xi^2}{2}\right)
{\bf K}(\xi)-{\bf E}(\xi)\right]\;, \label{m1m1}
\end{equation}
with ${\bf E}(\xi)$ as the complete elliptic
integral of second kind.  

From (\ref{ba}) follows:
\begin{equation}
h=\frac{2}{\pi}{\bf K}(\xi)+
\frac{4}{\xi^2}\left[
\frac{2}{\pi}{\bf E}(\xi)-1
\right]\;. \label{h}
\end{equation}
Therefore, from (\ref{qxi}):
\begin{equation}
q=\frac{1}{\xi}\left[
\frac{2}{\pi}{\bf E}(\xi)-1
\right]\;. \label{qxii}
\end{equation}
When $\xi \to 1^-$,  $t$, $u$, 
$b$ and $h$ all diverge, since
${\bf K}(\xi)$ has a 
logarithmic singularity in $\xi=1$.
Accordingly, we have the 
following asimptotic behaviour
\cite{Gradstheyn}:
\begin{equation}
t(\xi)\sim \frac{1}{\pi} \ln{\frac{8}{1-\xi}}
,\;\;\;\mbox{as}\;\;\;\xi 
\rightarrow 1^{-}. \label{limtxi}
\end{equation}
Note that for $\xi=1$, $q(1)=-(1-2/\pi)=-0.36338$.

In dimensions three  the value for $t(1)$ is given
analitically \cite{Hughes} as $t(1) = 
\frac{\sqrt{6}}{32\pi^3}
\Gamma(\frac{1}{24}) \\
\Gamma(\frac{5}{24})
\Gamma(\frac{7}{24}) 
\Gamma(\frac{11}{24}) 
= 1.516386$.
We numerically estimated the value of $b(1)$ as 
$b(1)=0.331$. Thus, from equation (\ref{ba}) 
$t(1)+4b(1)+h(1)=6u(1)$, and 
one finds $h(1)=0.2579$.
Therefore $q(1)=-0.209$.

\appendix

\noindent
{\bfit Diagonalization of \protect${\bf P}$}\\

${\bf P}$ is easily diagonalised. It has three different
eigenvalues (for $d=1$ only two, $\Lambda_1$ and 
$\Lambda_3$):
\begin{eqnarray}
\lambda_1=\Lambda_1 & = & h+2(d-1)b+t\;, \label{mat02} \\
\lambda_2=\lambda_3=...=\lambda_{d}=
\Lambda_2 & = & h-2b+t \;, \label{mat03} \\
\lambda_{d+1}=\lambda_{d+2}=...=
\lambda_{2d}=\Lambda_3 & = & h-t \;. \label{mat04} 
\end{eqnarray}
The corresponding system of orthonormal 
eigenvectors is:
\begin{eqnarray}
[\phi_1]^T = \frac{1}{\sqrt{2d}}
\left[ \begin{array}{ccccccc}
1 & \cdots & 1 & \Big| & 1 & \cdots & 1 \end{array}
\right]\;,  \label{mat05}
\end{eqnarray}
as the single eigenvector corresponding to $\Lambda_1$,
\begin{eqnarray}
[\phi_j]^T = \frac{1}{\sqrt{2j(j-1)}}
\Bigg[ -1\;0\;..\;0\;(j-1)\;-1\;..\;-1\!\!\!
&\Bigg|&\!\!\! -1\;0\;..\;0\;(j-1)\;-1
\;..\;-1 \Bigg] \nonumber \\
& & j\in \{2,3,...,d\}\;, \label{mat06} 
\end{eqnarray}
as the $d-1$ eigenvectors 
corresponding to $\Lambda_2$
(the small vertical line is 
between elements $d$ and $d+1$).
The elements $-1$ in $[\phi_j]^T$ 
are in the columns $1$, $d-j+3$, 
$d-j+4$, ..., $d-1$, $d$
 and $d+1$, $2d-j+3$, 
$2d-j+4$, ..., $2d-1$, $2d$,
 respectively, while
the elements $j-1$ are at the
 columns $d-j+2$ and $2d-j+2$,
respectively.
And, finally,
\begin{eqnarray}
[\phi_{i}]^T = \frac{1}{\sqrt{2}}\left[ 
\begin{array}{ccccccccccc}
0 &  \cdots & -1 &  \cdots & 0 & \Big| 
& 0 & \cdots & 1 & \cdots  & 0 \end{array}
\right],&& \nonumber \\
 \;\; i \in
\{ d+1,d+2,...,2d \}&& \label{mat07}
\end{eqnarray}
are the $d$ eigenvectors corresponding $\Lambda_3$,
with nonzero elements at columns $2d+1-i$ and $3d+1-i$.
It is  certainly valid:
\begin{equation}
 [\phi_i]^T [\phi_j] = 
 \delta_{i,j}\;,\;\;\; i,j 
 \in \{ 1,2,3,...,2d\}\;. \label{mat08}
\end{equation}
  
$\underline{d=1.}\;\;$ The eigenvalues with 
the corresponding eigenvectors are:
\begin{eqnarray}
\lambda_1=h+t=\frac{2}{\xi}u\;,\;\;
[\phi_1]=\frac{1}{\sqrt{2}}
\left[ \begin{array}{c}
1 \\ 1 \end{array} \right]\;,&& 
\mbox{and} \nonumber \\
\;\;\lambda_2=h-t=-\frac{2u}{\xi t}
\;,\;\;[\phi_2]=\frac{1}{\sqrt{2}}
\left[ \begin{array}{c}
-1 \\ 1 \end{array} \right] &&\label{eig1d}
\end{eqnarray}
with $t$ and $u$ given by Eqs.
 (\ref{pxi01d}) and (\ref{196}).
While $\lambda_1$ diverges 
as $\xi \to 1^-$, $\lambda_2$
is convergent ($u/t \to 1$).

$\underline{d=2.}\;\;$ In this case 
the eigenvalues are given by:

\begin{eqnarray}
\lambda_1 &=& h+2b+t=\frac{4}{\xi}
 u=\frac{4}{\xi^2} \left[
\frac{2}{\pi}{\bf K}(\xi)-1 \right]\;, \\
\lambda_2 &=& h-2b+t=\frac{4}{\xi^2} \left[
\frac{4}{\pi}{\bf E}(\xi)-1-
(1-\xi^2)\frac{2}{\pi}{\bf K}(\xi)\right]
\equiv \frac{4}{\xi} V \;, \\
\lambda_3 &=& \lambda_4 =h-t= \frac{4}{\xi^2}\left[  
\frac{2}{\pi}{\bf E}(\xi)-1\right]
\equiv \frac{4}{\xi} W\;,
\end{eqnarray}
and the corresponding eigenvectors by:
\begin{eqnarray}
[\phi_1]=\frac{1}{2} \left[ \begin{array}{r}
1 \\ 1 \\ 1 \\ 1 \end{array} \right],\;
[\phi_2]=\frac{1}{2} \left[ \begin{array}{r}
-1 \\ 1 \\ -1 \\ 1 \end{array} \right],\;
[\phi_3]=\frac{1}{\sqrt{2}} \left[ \begin{array}{r}
0 \\ -1 \\ 0 \\ 1 \end{array} \right],\;
[\phi_4]=\frac{1}{\sqrt{2}} \left[ \begin{array}{r}
-1 \\ 0 \\ 1 \\ 0 \end{array} \right]. &&\nonumber \\
 && \label{tirro}
\end{eqnarray}

\appendix

\noindent
{\bfit First and second moments}\\

 Using the expressions (\ref{Pikx})
and (\ref{Pi0x}) for the generating functions,
and performing the sums over
 ${\bf r}$ and $\mbox{\boldmath$\rho$}$, we have:
\begin{equation}
{\bf A}({\bf r_0};\xi) =
\frac{1-u/t}{1-\xi}
\sum_{k=1}^{\infty}
(p \xi)^k
\sum_{\mbox{{\scriptsize \boldmath$\nu_1$}}}^{({\bf 0})}
 ... \sum_{\mbox{\scriptsize \boldmath$\nu_k$}}^{({\bf 0})}
(\mbox{\boldmath$\nu_1$}+...+\mbox{\boldmath$\nu_k$})
P^{\dagger}(\mbox{\boldmath$\nu_{k}$}
\!\mid\!-\mbox{\boldmath$\mbox{\boldmath$\nu_{k-1}$}$};\xi) ...
P^{\dagger}(\mbox{\boldmath$
\nu_{1}$}\!\mid\!{\bf r_0};\xi)\;. \label{AC1}
\end{equation}
Using  (\ref{pdgz}) and the notation
(\ref{Mxi}), the $k=1$ term is:
\begin{equation}
\frac{1-u/t}{1-\xi} p \xi 
\sum_{\mbox{{\scriptsize \boldmath$\nu_1$}}}^{({\bf 0})}
\mbox{\boldmath$\nu_1$} P^{\dagger}
(\mbox{\boldmath$\nu_{1}$}\!\mid\!{\bf r_0};\xi)=
\frac{1-u/t}{1-\xi}\;{\bf M}({\bf r_0};\xi)\;.
\end{equation}
Defining:
\begin{equation}
{\bf A_1}({\bf r_0};\xi) \equiv
\sum_{k=2}^{\infty}
(p \xi)^k
\sum_{\mbox{{\scriptsize \boldmath$\nu_1$}}}^{({\bf 0})}
 ... \sum_{\mbox{\scriptsize \boldmath$\nu_k$}}^{({\bf 0})}
(\mbox{\boldmath$\nu_1$}+...+\mbox{\boldmath$
\mbox{\boldmath$\nu_{k-1}$}$})
P^{\dagger}(\mbox{\boldmath$\nu_{k}$}\!\mid\!
-\mbox{\boldmath$\mbox{\boldmath$\nu_{k-1}$}$};\xi) ...
P^{\dagger}(\mbox{\boldmath$\nu_{1}$}\!\mid\!
{\bf r_0};\xi)\;, \label{AC11}
\end{equation}
and 
\begin{equation}
{\bf A_2}({\bf r_0};\xi) \equiv
\sum_{k=2}^{\infty}
(p \xi)^k
\sum_{\mbox{{\scriptsize \boldmath$\nu_1$}}}^{({\bf 0})}
 ... \sum_{\mbox{\scriptsize \boldmath$\nu_k$}}^{({\bf 0})}
\mbox{\boldmath$\nu_{k}$}
P^{\dagger}(\mbox{\boldmath$\nu_{k}$}\!\mid\!
-\mbox{\boldmath$\mbox{\boldmath$\nu_{k-1}$}$};\xi) ...
P^{\dagger}(\mbox{\boldmath$\nu_{1}$}\!\mid\!
{\bf r_0};\xi)\;, \label{AC12}
\end{equation}
(\ref{AC1}) is rewritten as:
\begin{equation}
{\bf A}({\bf r_0};\xi) =
\frac{1-u/t}{1-\xi}\;
\left[ {\bf M}({\bf r_0};\xi) +
{\bf A_1}({\bf r_0};\xi) +{\bf A_2}({\bf r_0};\xi)  
\right]. \label{suma}
\end{equation}
It is easy to see that:
\begin{equation}
{\bf A_1}({\bf r_0};\xi) =
\frac{u}{t}\;
\frac{1-\xi}{1-u/t}\;{\bf A}({\bf r_0};\xi)\;. \label{aa1}
\end{equation}
With the aid of formulas (\ref{part4}) and (\ref{Mxi})
for ${\bf A_2}({\bf r_0};\xi)$ we obtain:
\begin{equation}
{\bf A_2}({\bf r_0};\xi) =
\frac{q}{1-q}\;{\bf M}({\bf r_0};\xi)\;. \label{aa2}
\end{equation}
Inserting (\ref{aa1}) and (\ref{aa2})
into (\ref{suma}), (\ref{Axi}) is gained.

 Proceeding similarly as for the first 
 moment, the
 generating function for the second 
 moment becomes:
\begin{equation}
S({\bf r_0};\xi)=
\frac{1-u/t}{1-\xi}
\sum_{k=1}^{\infty}
(p \xi)^k
\sum_{\mbox{{\scriptsize \boldmath$\nu_1$}}}^{({\bf 0})}
 ... \sum_{\mbox{\scriptsize \boldmath$\nu_k$}}^{({\bf 0})}
(\mbox{\boldmath$\nu_1$}+...+\mbox{\boldmath$\nu_k$})^2
P^{\dagger}(\mbox{\boldmath$\nu_{k}$}\!\mid\!
-\mbox{\boldmath$\mbox{\boldmath$\nu_{k-1}$}$};\xi) ...
P^{\dagger}(\mbox{\boldmath$\nu_{1}$}
\!\mid\!{\bf r_0};\xi). \label{SC1}
\end{equation}
Separating the term $k=1$ in (\ref{SC1}) and using
the fact that $(\mbox{\boldmath$\nu_1$}+...
+\mbox{\boldmath$\nu_k$})^2=
(\mbox{\boldmath$\nu_1$}+...+\mbox{\boldmath$
\mbox{\boldmath$\nu_{k-1}$}$})^2+2(\mbox{\boldmath$\nu_1$}+...
+\mbox{\boldmath$\mbox{\boldmath$\nu_{k-1}$}$})+1$ 
($\mbox{\boldmath$\nu_i$}^2=1$, $\forall i$), 
$S({\bf r_0};\xi)$ is rewritten
as:
\begin{equation}
S({\bf r_0};\xi)=
\frac{1-u/t}{1-\xi}
\left[\frac{P({\bf 0}\!\mid\!{\bf r_0};\xi)}{t}+
S_1({\bf r_0};\xi)+2 S_2({\bf r_0};\xi)+S_3({\bf r_0};\xi) 
\right], \label{sums}
\end{equation}
where
\begin{equation}
S_1({\bf r_0};\xi)=
\sum_{k=2}^{\infty}
(p \xi)^k
\sum_{\mbox{{\scriptsize 
\boldmath$\nu_1$}}}^{({\bf 0})}
 ... \sum_{\mbox{\scriptsize \boldmath$\nu_k$}}^{({\bf 0})}
(\mbox{\boldmath$\nu_1$}+...+
\mbox{\boldmath$\mbox{\boldmath$\nu_{k-1}$}$})^2
P^{\dagger}(\mbox{\boldmath$\nu_{k}$}\!\mid\!
-\mbox{\boldmath$\mbox{\boldmath$\nu_{k-1}$}$};\xi) ...
P^{\dagger}(\mbox{\boldmath$\nu_{1}$}\!\mid\!
{\bf r_0};\xi), \label{S1C1}
\end{equation}
\begin{equation}
S_2({\bf r_0};\xi)=
\sum_{k=2}^{\infty}
(p \xi)^k
\sum_{\mbox{{\scriptsize \boldmath$\nu_1$}}}^{({\bf 0})}
 ... \sum_{\mbox{\scriptsize \boldmath$\nu_k$}}^{({\bf 0})}
(\mbox{\boldmath$\nu_1$}+...+
\mbox{\boldmath$\mbox{\boldmath$
\nu_{k-1}$}$})\mbox{\boldmath$\nu_k$}
P^{\dagger}(\mbox{\boldmath$\nu_{k}$}
\!\mid\!-\mbox{\boldmath$
\mbox{\boldmath$\nu_{k-1}$}$};\xi) ...
P^{\dagger}(\mbox{\boldmath$\nu_{1}$}
\!\mid\!{\bf r_0};\xi)\;, \label{S2C1}
\end{equation}
\begin{equation}
S_3({\bf r_0};\xi)=
\sum_{k=2}^{\infty}
(p \xi)^k
\sum_{\mbox{{\scriptsize \boldmath$\nu_1$}}}^{({\bf 0})}
 ... \sum_{\mbox{\scriptsize \boldmath$\nu_k$}}^{({\bf 0})}
P^{\dagger}(\mbox{\boldmath$\nu_{k}$}\!\mid\!
-\mbox{\boldmath$\mbox{\boldmath$\nu_{k-1}$}$};\xi) ...
P^{\dagger}(\mbox{\boldmath$
\nu_{1}$}\!\mid\!{\bf r_0};\xi)\;, \label{S3C1}
\end{equation}
In virtue of (\ref{part3}) and  
(\ref{part4}) is not hard to see that:
\begin{equation}
S_1({\bf r_0};\xi)=
\frac{u}{t}
\frac{1-\xi}{1-u/t}S({\bf r_0};\xi)\;,\;\;\;
S_2({\bf r_0};\xi)=
\frac{q}{1-q}\; 
\frac{1}{1-u/t}\; 
\frac{P({\bf 0}\!\mid\!{\bf r_0};\xi)}{t}\;, \label{sa2}
\end{equation}
and
\begin{equation}
S_3({\bf r_0};\xi)=
\frac{u}{t}\; 
\frac{1}{1-u/t}\; 
\frac{P({\bf 0}\!\mid\!{\bf r_0};\xi)}{t}\;, \label{sa3}
\end{equation}
Inserting these into (\ref{sums}),
(\ref{Sxi}) follows.

\vfill

\end{document}